\documentclass[12pt]{article}

\usepackage{latexsym}
\usepackage{amsmath}
\usepackage{verbatim}
\usepackage{amssymb}
\usepackage{multirow}
\usepackage[T1]{fontenc}
\usepackage{charter}
\usepackage{physics}
\usepackage{fancyhdr}
\usepackage{hyperref}
\usepackage{lastpage}
\usepackage{xcolor}
\usepackage{tikz}
\usepackage{wasysym}
\usepackage[
top    = 2.75cm,
bottom = 2.75cm, left=20mm,right=20mm]{geometry}
\usepackage{slashed}
\usepackage{nicefrac}
\usepackage{mathrsfs}

\definecolor{brightpurple}{RGB}{138,43,226}   %
\definecolor{brightmagenta}{RGB}{255,0,255}   
\definecolor{brightviolet}{RGB}{148,0,211}  
\definecolor{arXiv}{RGB}{255,0,255} 
\hypersetup{
    colorlinks=true,
    linkcolor=brightpurple,     % internal links (TOC, section titles)
    citecolor=brightmagenta,    % citations
    urlcolor=brightviolet        % external URLs
}
            
\numberwithin{equation}{section}

\begin{document}
\title{Ultraviolet completion of Starobinsky inflation}

\newcommand{\sns}{\textcolor{brightpurple}{\blacklozenge}}
\newcommand{\infn}{\textcolor{brightpurple}{\blacksquare}}
\newcommand{\ias}{\textcolor{brightpurple}{\blacktriangleleft}}
\newcommand{\lpthe}{\textcolor{brightpurple}{\blacktriangleright}}
\newcommand{\aoath}{\textcolor{brightpurple}{\bigstar}}
\newcommand{\tamu}{\textcolor{brightpurple}{\bullet}}

\newcommand{\D}{\texttt{D}}
\newcommand{\F}{\texttt{F}}

\author{%
Ignatios Antoniadis\footnote{antoniad@lpthe.jussieu.fr} $^{\ias,\lpthe}$,
Chrysoula Markou\footnote{chrysoula.markou@sns.it} $^{\sns,\infn}$ and %
Dimitri V. Nanopoulos\footnote{dimitri@physics.tamu.edu} $^{\aoath,\tamu}$
}

\date{} % remove date
\maketitle

\begin{center}
\begin{minipage}{0.9\textwidth}
\centering
\textit{%
$^{\ias}$School of Natural Sciences, Institute for Advanced Study, Princeton NJ 08540, USA \\
$^{\lpthe}$Laboratoire de Physique Th\'eorique et Hautes Energies -- LPTHE, \\ Sorbonne Universit\'e, CNRS, 4 Place Jussieu, 75005 Paris, France \\
$^{\sns}$Scuola Normale Superiore, Piazza dei Cavalieri 7, 56126 Pisa, Italy \\
$^{\infn}$INFN, Sezione di Pisa, Largo Pontecorvo 3, 56127, Pisa, Italy \\
$^{\aoath}$Academy of Athens, Division of Natural Sciences, Athens 10679, Greece\\
$^{\tamu}$George P. and Cynthia W. Mitchell Institute for Fundamental Physics and Astronomy, Texas A\&M University, College Station, TX 77843, USA
}
\end{minipage}
\end{center}

\begin{abstract}
We construct an ${\cal N}=1$ supergravity action whose bosonic part contains an arbitrary function of the scalar curvature, the so-called $F(R)$ gravity. As in $R+R^2$ supergravity, it can be described in terms of two chiral superfields of no-scale supergravity: one contains the scalaron which plays the role of the Starobinsky inflaton and the other contains the goldstone fermion of spontaneously broken supersymmetry during the inflation plateau. Its (complex) scalar component acquires a non-tachyonic mass in the presence of the string dilaton and can be set to zero, together with the pseudoscalar partner of the scalaron, so that the scalar potential is reduced to the one of $F(R)$ gravity. In a perturbative expansion in powers of $R$, one obtains a small deformation of the Starobinsky cosmological model that solves the problem of initial conditions within the validity of the effective field theory, below the scale of tower of states predicted by the swampland distance conjecture. We also show that a particular example of an underlying microscopic theory with such properties is provided by a four-dimensional heterotic string model containing the Standard Model of particle physics.
\end{abstract}
 \newpage
\tableofcontents

\newpage
%%%%%%%%%%%%%%%%%%%%%%%%%%%%%%%
\section{Introduction} \label{sec:intro}
%%%%%%%%%%%%%%%%%%%%%%%%%%%%%%%
Although there is no actual experimental evidence for low energy supersymmetry, it may still be hidden around and within experimental search in present and future particle colliders, while there are strong theoretical reasons that it should be part of the underlying fundamental theory of gravity, such as string theory, and may play important an role at higher energy scales and particularly in cosmology~\cite{Ellis:1982ed,Ellis:1982ws}. In this context, the Starobinsky model of inflation~\cite{Starobinsky:1980te}, which is very attractive because of its simplicity and very good description of the observational data~\cite{Planck:2018jri, Planck:2018vyg, Tristram:2021tvh}, is known to have a minimal supersymmetric extension as ordinary ${\cal N}=1$ supergravity coupled to two chiral multiplets~\cite{Cecotti:1987sa}. One of the two contains the scalar degree of freedom of $R^2$ gravity that plays the role of the inflaton which is complexified due to supersymmetry, and the other contains the goldstino of spontaneously broken supersymmetry during inflation~\cite{Ellis:2013xoa, Ferrara:2013wka}. When its (complex) scalar component (sgoldstino) is set to zero, together with the pseudoscalar partner of the inflaton, the scalar potential is reduced to the one of the Starobinsky model.

Despite its phenomenological success, Starobinsky inflation suffers from two theoretical problems: its validity as an effective field theory and the problem of initial conditions along the flat region of the scalar potential that extends to infinity~\cite{Antoniadis:2024ypf, Antoniadis:2025pfa}. Moreover, its supersymmetric extension suffers from an instability during inflation because the sgoldstino becomes tachyonic. 
\begin{itemize}
\item 
The problem of effective field theory (EFT) validity has two aspects: (i) the effect of higher order terms, in particular powers of the curvature scalar, and (ii) the effect of trans-Planckian excursions of the inflaton, typical to models of large field inflation. The first aspect can be studied by adding an $R^3$ term in the action~\cite{Berkin:1990nu, Gialamas:2025ofz, Gialamas:2025thp}, which can be generalised to an arbitrary function $F(R)$~\cite{Buchdahl:1970ldb, Huang:2013hsb, Qiu:2025uot}
(for reviews see e.g.~\cite{Sotiriou:2008rp,DeFelice:2010aj}). The second aspect can be addressed in the framework of the Swampland program~\cite{Vafa:2005ui}, using the large distance conjecture~\cite{Ooguri:2006in} that implies a tower of light states at a mass scale which is exponentially small in the proper distance with an order one exponent in Planck units~\cite{Agrawal:2018own, Scalisi:2018eaz, Lust:2023zql}. It turns out that this tower of states is a Kaluza-Klein (KK) tower associated to the decompactification of extra dimensions~\cite{Lee:2019wij, Antoniadis:2024ypf}. Although the computation of the mass of the tower was done for a string model which is supersymmetric and shares the properties of the Starobinsky model (same superpotential and same K\"ahler potential for vanishing sgoldstino)~\cite{Antoniadis:2020txn}, it was estimated for the latter using the effective supergravity~\cite{Antoniadis:2024ypf}. Actually the order of magnitude of the result can be found by applying simply the swampland distance conjecture with a coefficient of the distance in the exponent equal to unity, with the distance being the variation of the canonically normalised inflaton field.
Since a KK tower implies that the graviton has massive spin-2 excitations, the compactification scale should be higher that the Hubble scale of slow-roll inflation $H$ due to unitarity (Higuchi bound)~\cite{Higuchi:1986py}, leading to an upper bound of the inflaton excursion, and thus on the number of e-folds of inflation~\cite{Antoniadis:2024ypf}. 
\item 
In a model with flat asymptotic potential, the initial condition defining the beginning of inflation is naturally far in the field space and thus free of any tuning, in contrast to other inflation models. On the other hand, this would imply a long period of inflation which although unobservable today it would be observable in the future with longer and longer wave lengths entering the horizon. As mentioned above, if inflation has lasted sufficiently long, EFT would break down when the mass of the tower becomes lower than $H$ and moreover unitarity would be violated. Thus, the problem of initial conditions comes back as a question of why the start of inflation should be at a narrow region below the critical value of the inflaton associated to the mass scale of the KK tower. It has been observed that a small deformation of the Starobisnky model can lead to a sharp rise of the scalar potential~\cite{Ellis:2013xoa,Antoniadis:2020txn, Gialamas:2025ofz}, reducing the flat asymptote towards infinity to a finite region and thus addressing the problem of initial conditions~\cite{Antoniadis:2025pfa}. 
\item
Finally, one way to address the instability of the supersymmetric $R+R^2$ model is by modifying the K\"ahler potential of the chiral fields in the equivalent ${\cal N}=1$ supergravity description~\cite{Ellis:1984bs, Ellis:2013nxa, Kallosh:2013lkr, Kallosh:2013maa} at the expense of loosing the nice geometric interpretation of the model, or to make the supersymmetry nonlinearly realised by imposing a nilpotent constraint that eliminates the scalar component of the goldstino multiplet~\cite{Antoniadis:2014oya, Antoniadis:2015ala}. Alternatively, it was recently pointed out that inclusion of the string dilaton turns the sgoldstino non tachyonic in the inflation region of the scalar potential, eliminating the instability~\cite{Antoniadis:2020txn, Antoniadis:2024ypf}.
\end{itemize}
 
In this work, we address all the above questions by providing a simple ${\cal N}=1$ supergravity framework of no-scale type that generalises Starobinsky $R+R^2$ supergravity to an arbitrary real function $F(R)$ and we identify a string theory embedding within four dimensional (4D) heteroric superstrings. In a non-supersymmetric context, it is known that $F(R)$ gravity introduces a scalar degree of freedom, besides the massless graviton, which reduces to the so-called scalaron of $R+R^2$~\cite{Stelle:1976gc, Stelle:1977ry, Sotiriou:2008rp}. Unlike other four-derivative terms, the scalaron has positive norm and is non-tachyonic when the sign of $R^2$ is positive in the Euclidean action~\cite{Stelle:1976gc, Barth:1983hb}. Moreover, it can be described as ordinary Einstein gravity coupled to a scalar field~\cite{Whitt:1984pd} with a positive definite potential behaving as a constant at infinity, and thus providing a simple model of (large field) inflation which is in remarkable agreement with cosmological observations of the cosmic microwave background (CMB) anisotropies. Motivated by recent CMB data~\cite{AtacamaCosmologyTelescope:2025blo, DESI:2024mwx}, signalling potential deviations from the perfect agreement, as well as by theoretical considerations on the problem of initial conditions and the EFT validity, possible deformations of the Starobinsky cosmological model were recently considered~\cite{Antoniadis:2025pfa}. In particular, the effect of an $R^3$ term was analysed and it was shown that when its coupling is sufficiently small and of a particular sign, the scalar potential acquires a barrier preventing the inflaton from having large field excursions above a certain value, while it gives a better fit to experimental data~\cite{Gialamas:2025ofz, Gialamas:2025thp}.

In a supersymmetric context, one would naively expect that the scalaron would be complexified as part of a chiral multiplet coupled to ${\cal N}=1$ supergravity. However, a difficulty to supersymmetrise $R^2$ or in general $F(R)$ is due to the fact that the scalar curvature is in an upper component of a chiral superfield ${\cal R}$, and thus the Einstein-Hilbert action is obtained by a chiral integral of ${\cal R}$ over half of the superspace (F-term), while its square ${\cal R}^2$ does not contain $R^2$ in its upper component. In fact, an $\mathcal{N}=1$ $f(\mathcal{R})$ supergravity Lagrangian was proposed and studied in~\cite{Gates:2009hu, Ketov:2009wc, Ketov:2010qz}, but the corresponding scalar potential does not reproduce the Starobinsky model and is insufficient for describing the data~\cite{Ketov:2013dfa, Ferrara:2013wka}. On the other hand, a supersymmetric action leading to an $R^2$ bosonic term is obtained by a D-term ${\cal R}\bar{\cal R}$ integrated over the full superspace, which can be linearised in terms of two chiral superfields  coupled to $\mathcal{N}=1$ supergravity~\cite{Cecotti:1987sa} which is of no-scale type~\cite{Cremmer:1983bf, Lahanas:1986uc}. The additional chiral multiplet contains the sgoldstino which becomes tachyonic at large inflaton values during inflation. However, in the presence of the string dilaton, the sgolstino's mass$^2$ turns positive everywhere in the inflaton field space and can be set to zero, leading to the Starobinsky scalar potential in the real field direction~\cite{Antoniadis:2024ypf}. In this work, we generalise the above supergravity construction and show that these results continue to hold for any function $F(R)$.

A string realisation of Starobinsky supergravity was provided~\cite{Antoniadis:2020txn} within the framework of the free-fermionic construction of 4D heterotic superstrings~\cite{Antoniadis:1986rn}, which contains the flipped $SU(5)\times U(1)$ grand unified gauge group and three chiral generations of quarks and leptons leading to realistic particle phenomenology~\cite{Antoniadis:1989zy, Antoniadis:2021rfm}. During inflation, the observable gauge group is unbroken, while the inflaton sector consists of two singlet chiral superfields which play the role of those that linearise the Starobinsky model~\cite{Antoniadis:2020txn}. Their superpotential is generated at 6th and 8th order in (the inverse string tension) $\alpha'$-expansion away from the free-fermionic vacuum, driven by the cancellation of a $U(1)$ anomaly, so that all moduli are fixed with the exception of the string dilaton, for which a separate stabilisation mechanism must be implemented. The scale of inflation is then fixed within the energy range required by observations, while the sgoldstino is massive (non-tachyonic) and can be set to zero. The superpotential is a sum of two contributions of the same order with a relative coefficient $\lambda$, such that $\lambda=1$ corresponds to the (supesymmetric) Starobinsky superpotential, while a slightly different value with $\lambda\lesssim 1$ produces a sharp rise of the scalar potential preventing large field excursions of the inflaton. Here, we show that the correction proportional to $1-\lambda$ corresponds to an $R^3$ deformation in the geometric formulation of the action. Moreover, possible higher-order corrections to the string superpotential can be mapped to a general $F(R)$ gravity deforming the Starobinsky model by higher powers of the scalar curvature.  

The outline of this paper is the following. In Section~\ref{sec:bosonic}, we give an overview of the Starobinsky $R+R^2$ gravity model, the addition of an $R^3$ term and its generalisation to $F(R)$ gravity. In Section~\ref{sec:sugra}, we  discuss the supergravity embedding. We first present an overview of the Starobinsky supergravity (subsection~\ref{subsec:Starsugra}), we then continue with the supersymmetrisation of $R^3$ (subsection~\ref{subsec:mod}) and we proceed with constructing the general ${\cal N}=1$ $f(R)$ supergravity containing in its bosonic sector the $f(R)$ gravity (subsection~\ref{subsec:f(R)}). Section~\ref{sec:string} contains the superstring embedding. We first present a short overview of the ANR model~\cite{Antoniadis:1989zy, Antoniadis:2020txn, Antoniadis:2021rfm}  (subsection~\ref{subsec:su(5)}) and then we describe the string embedding of $f(R)$ supergravity (subsection~\ref{subsec:embed}). Our conclusions are summarised in Section~\ref{sec:concl}. Finally, the Appendix~\ref{app:ids} contains some basic elements of $\mathcal{N}=1$ supergravity which is used in Section~\ref{sec:sugra}.

\paragraph{Conventions.} We work in the metric formulation of gravity and with the mostly plus metric signature. For most of this work the reduced Planck mass is set to $1$, namely $M_{\textrm{Pl}}=1$. The real and imaginary parts of a superfield $Y$ will be denoted by $Y_{\mathtt R}$ and $Y_{\mathtt I}$ respectively.

%%%%%%%%%%%%%%%%%%%%%%%%%%%%%%%
\section{\texorpdfstring{Starobinsky gravity, the \ensuremath{R^3}  modification and \ensuremath{F(R)} gravity}{Starobinsky gravity, the R3 modification and f(R) gravity}} \label{sec:bosonic}
%%%%%%%%%%%%%%%%%%%%%%%%%%%%%%%

The Lagrangian of the Starobinsky model of modified gravity \cite{Starobinsky:1980te} reads
\begin{align} \label{eq:starob}
    \mathcal{L} = \tfrac12 \sqrt{-g} \big( R + \alpha R^2  \big)\,,
\end{align}
where $\alpha=\nicefrac{1}{6 m^2}>0$, $m$ being an a priori free parameter with dimensions of mass. The Lagrangian \eqref{eq:starob}, being, like General Relativity (GR), ghost--free but non--renormalisable in four dimensions \cite{Stelle:1976gc,Stelle:1977ry, Barth:1983hb}, is dual to canonical gravity coupled to a scalar field  \cite{Whitt:1984pd}. In particular, upon introducing two scalar fields, $\chi$ and $\phi$, the Lagrangian \eqref{eq:starob} may be written in the dual form
\begin{align} \label{eq:starob1}
    \mathcal{L} = \tfrac12 \sqrt{-g} \big[  R + 2\, \phi (R-\chi) + \alpha \chi^2  \big]\,,
\end{align}
where $\phi$ acts as a Lagrange multiplier in order to recover \eqref{eq:starob}. Integrating out, however, the field $\chi$, brings the Lagrangian \eqref{eq:starob1} to a Brans--Dicke form in the Jordan frame,
\begin{align} \label{eq:starob2}
    \mathcal{L} = \tfrac12 \sqrt{-g} \bigg[ ( 1 + 2\, \phi ) R - \frac{1}{\alpha} \phi^2  \bigg]\,.
\end{align}
After a Weyl rescaling of the metric and a scalar field redefinition to make it canonically normalised, namely the field replacements
\begin{align} \label{eq:wfr}
 g_{\mu \nu} \rightarrow (1+2 \phi )^{-1} g_{\mu \nu} \,, \quad    \phi \rightarrow \tfrac12 \bigg( e^{\sqrt{\tfrac23}\phi}-1 \bigg) \,,
\end{align}
the Lagrangian \eqref{eq:starob2} is brought to its form in the Einstein frame, namely 
\begin{align} \label{eq:lagr_einst}
        \mathcal{L} =\sqrt{-g} \Big( \tfrac12  R - \tfrac12  \partial_\mu \phi \partial^\mu \phi -V(\phi) \Big)\,,
\end{align}
where $V\equiv V_{\textrm{Star}}$ is the Starobinsky scalar potential,
\begin{align} \label{eq:starpot}
    V_{\textrm{Star}}(\phi) =  \tfrac34 m^2 \bigg(1-e^{-\sqrt{\tfrac23}\phi}\bigg)^2 \,.
\end{align}

The Starobinsky potential \eqref{eq:starpot} is defined for $\phi>-1/2$ where the rescaling \eqref{eq:wfr} becomes singular and it
possesses an asymptote at large positive field values, approaching exponentially a constant value for $\phi \gg 1$; on the other hand, its minimum is at $\phi=0$ with zero vacuum energy, with the inflaton mass given by $m$, describing the spectrum around flat space in the geometric formulation. The parameter $m$ defines the scale of inflation with the Hubble parameter being $H=m/2$, while the normalisation of the CMB spectrum imposes $m \sim 10^{-5} M_{\textrm{Pl}}$, namely it forces $m$ to be  much larger than the electroweak scale but much smaller than the Planck mass. The values of the spectral index $n_s\simeq 0.9649$ and of the tensor--to--scalar ratio $r\simeq 0.0035$ that are predicted by the Starobinsky model are consistent with PLANCK data \cite{Planck:2018jri,Planck:2018vyg,Tristram:2021tvh}. They are computed at the horizon exit corresponding to $\phi_\star\simeq 5.35$, around 55 e-folds before the end of inflation corresponding to $\phi_\text{end}\simeq 0.62$. 

It follows that the inflaton takes a trans--Planckian excursion ($\delta\phi > M_{\textrm{Pl}}$) during inflation, implying that the effective field theory is no longer valid at very high energies. %, up to $M_{\textrm{Pl}}$. 
Actually, the EFT should remain valid (at least) up to the inflationary Hubble scale $H$.\footnote{For a discussion of these considerations, although in a different model, see~\cite{Cicoli:2017axo}.} On the other hand,
the swampland distance conjecture predicts the appearance of a tower of light states \cite{Ooguri:2006in} at a scale $m_{\textrm{tower}}\sim e^{-c\delta\phi}$ exponentially small in the proper distance $\delta\phi$ with an exponent $c$ of order unity in Planck units. The tower can be analysed within string supersymmetric compactifications and is associated with the decompactification of extra dimensions~\cite{Antoniadis:2024ypf}. Therefore, it includes massive spin--$2$ KK graviton excitations, whose mass should respect the Higuchi bound \cite{Higuchi:1986py}, $m_\text{spin-2}>\sqrt{2}H$, thereby implying a breakdown of the effective field theory at the scale of the tower $m_{\textrm{tower}}\sim m_\text{spin-2}$ and restricting the total number of e-folds from the start of inflation by imposing $\phi_\text{start}\lesssim 10$. 
This can be easily seen by imposing $e^{-\delta\phi}\gtrsim H/M_{\textrm{Pl}}\sim 10^{-5}$.
Thus, the initial condition of the inflaton should be within a factor of 2 higher than its value at the horizon exit which looks unnatural since the plateau of the scalar potential extends to infinity.

Now let us turn to the generic case of $F(R)$ gravity (see for example the reviews \cite{Sotiriou:2008rp,DeFelice:2010aj}), namely the extension of the Einstein--Hilbert action to the Lagrangian
\begin{align} \label{eq:fR}
    \mathcal{L} = \tfrac12 \sqrt{-g}  F(R)   \,,
\end{align}
recalling that this class of theories appears to be the only higher--order extension of gravity that may not excite the pathological Ostrogradski ghost \cite{Woodard:2006nt}. If $F(R)$ is ``$R$--regular'', namely if its second derivative $F''$ is defined and the differentiability condition $F'' \neq 0$ holds  (with the singular case corresponding to GR),  the Lagrangian \eqref{eq:fR} may be written in the dual form
\begin{align} \label{eq:f1}
    \mathcal{L} = \tfrac12 \sqrt{-g} \big[F(\chi) + F'(\chi)(R-\chi)  \big] \,,
\end{align}
where the scalar field $\chi$ acts as a Lagrange multiplier in order to recover \eqref{eq:fR}. Moreover, by setting $\varphi:= F'(\chi)$, namely treating the first derivative of $F$ as a (scalar) degree of freedom (and assuming that $F'$ is invertible and yields $\chi = \chi(\varphi)$\footnote{The necessary condition for the equivalence of the Lagrangians \eqref{eq:fR} and \eqref{eq:f1} is in fact simply that $F'$ be continuous and one-to-one, while the condition $F''\neq 0$ is sufficient but not necessary \cite{Olmo:2006eh}.}), the Lagrangian is brought to a Brans--Dicke form in the Jordan frame,
\begin{align} \label{eq:f2}
    \mathcal{L} = \tfrac12 \sqrt{-g} \big[\varphi R -  \varphi \chi (\varphi) + F\big(\chi(\varphi)\big)  \big]
   \,;\quad \varphi := F'(\chi) \,.
\end{align}
The Weyl rescaling and canonically normalised field redefinition
\begin{align} \label{eq:wfr2}
    g_{\mu \nu} \rightarrow \frac{1}{\varphi} \, g_{\mu \nu} \,,\quad \varphi = e^{\sqrt{\tfrac23}\phi},
\end{align}
bring the Lagrangian \eqref{eq:f2} to the form \eqref{eq:lagr_einst} in the Einstein frame, with a potential that reads 
\begin{align}\label{f(R)potential}
    V(\phi)= \frac{\chi(\phi)F'\big( \chi(\phi) \big) -F \big( \chi(\phi) \big)  }{2\Big(F'\big(\chi (\phi) \big)\Big)^2} \,.
\end{align}
To avoid ghosts and tachyons, the function $F$ has to further satisfy the conditions (if $F''$ is not identically zero) \cite{Dolgov:2003px,Starobinsky:2007hu}
\begin{align} \label{eq:conditionsf}
    F'(R)>0\,,\quad F''(R)>0\,.
\end{align}
Note also that positivity of the scalar potential \eqref{f(R)potential} requires the condition
\begin{align} \label{positivityV}
\left(\frac{F}{\chi}\right)'>0\,
\end{align}
which may not be necessary in the bosonic theory but will be required by unitarity in its supersymmetrisation, we will see in Section~\ref{subsec:f(R)}.

%The special choice
%\begin{align}
%    F(R) = R + \epsilon \tilde{f}(R)\,,
%\end{align}
%where $\epsilon$ is a ``small'' dimensionful parameter and $\tilde{f}(R)$ a function of $R$ and its powers can then be made. Particular such cases are 
A particular case is for $F$ being a cubic polynomial corresponding to the Starobinsky model and its modification  by a term cubic in the Ricci scalar. This is described by the Lagrangian
\begin{align} \label{eq:modif}
    \mathcal{L} = \tfrac12 \sqrt{-g} \big( R + \alpha R^2  + \beta R^3 \big)\,,
\end{align}
where $\beta$ is an a priori arbitrary parameter with mass dimension equal to $-4$. The model \eqref{eq:modif} was investigated in \cite{Berkin:1990nu} and more recently in \cite{Gialamas:2025ofz, Gialamas:2025thp}, while its $F(R)$ extension in \cite{Huang:2013hsb, Qiu:2025uot}. In this case, inverting $\varphi= F'(\chi) $ and redefining $\varphi$ as in \eqref{eq:wfr2} yields two solutions
\begin{align}
   \chi^{\pm} =  \frac{-\alpha\pm\sqrt{\alpha ^2  -3 \beta +3 \beta e^{\sqrt{\frac{2}{3}} \phi } } }{3 \beta }\,,
\end{align}
of which only the solution $\chi^+$ respects the condition $F''>0$. In that case, the potential  reads 
\begin{align} \label{eq:vstarmod}
    V(\phi) = \frac{m^2 }{144 \beta ^2} e^{-2 \sqrt{\frac{2}{3}} \phi } \left( \sqrt{12 \beta  \left(e^{\sqrt{\frac{2}{3}} \phi }-1\right)+1}-1\right)^2 \left( 2\sqrt{12 \beta  \left(e^{\sqrt{\frac{2}{3}} \phi }-1\right)+1}+1\right) \,,
\end{align}
where we have rescaled the parameter $\beta$ via the replacement $\beta \rightarrow \tfrac{1}{9m^4} \beta$ such that $\beta$ be now dimensionless. 
The condition $F'>0$ implies that $\phi$ is positive, while the reality of the argument of the square root implies that either $\beta\ge 0$, or $|\beta|<1/12$ for $\beta$ negative, in which case $\phi$ is bounded and the potential rises sharply restricting large excursions of the inflation; see discussion in Section~\ref{subsec:mod} and Figure~\ref{fig:potR3}.
This is illustrated for $|\beta| \ll 1$, where the potential \eqref{eq:vstarmod} can be expanded as 
\begin{align}
    V(\phi) =   V_{\textrm{Star}}(\phi) - \tfrac{3}{2} \beta m^2 \,  e^{ \sqrt{\frac{2}{3}} \phi } \left(1- e^{-\sqrt{\frac{2}{3}} \phi }\right)^3 + \mathcal{O}(\beta^2)\,,
\end{align}
namely it is the Starobinsky potential with a correction linear in $\beta$ that rises for $\beta<0$.

%%%%%%%%%%%%%%%%%%%%%%%%%%%%%%%
\section{Supergravity extension and inflation} \label{sec:sugra}
%%%%%%%%%%%%%%%%%%%%%%%%%%%%%%%

%%%%%%%%%%%%%%%%%%%%%%%%%%%%%%%
\subsection{A brief review of Starobinsky no--scale supergravity} \label{subsec:Starsugra}
%%%%%%%%%%%%%%%%%%%%%%%%%%%%%%%

The Starobinsky model \eqref{eq:starob} has a simple embedding in $\mathcal{N}=1$ supergravity \cite{Cecotti:1987sa,Ferrara:2013wka} in the old--minimal version formalism \cite{Freedman:2012zz}, that is given by the  Lagrangian 
\begin{align} \label{eq:starsugra}
\mathcal{L} =  3 \bigg[\bigg(-1+ 6 \alpha \frac{\mathcal{R}}{S_0}\frac{\bar{\mathcal{R}}}{\bar{S}_0}   \bigg) S_0 \bar{S}_0  \bigg]_\D  \,,
\end{align}
where $S_0$ is the chiral compensator superfield and $\mathcal{R}$ the chiral curvature superfield. The latter contains the Ricci scalar $R$ in its highest (F-)component, such that the $R^2$ term appears due to the $\mathcal{R} \bar{\mathcal{R}}$ term in the $\D$--density \eqref{eq:starsugra}, leading to the bosonic Lagrangian~\eqref{eq:starob}. By introducing \textit{two} chiral superfields, $T$ and $C$,  the Lagrangian \eqref{eq:starsugra} can be written in the dual form
\begin{align} \label{eq:starsugra2}
    \mathcal{L} =  3 \Big[\big(-1+ 6 \alpha \, C \bar{C} \big) \, S_0 \bar{S}_0  \Big]_\D  +  
   3 \bigg( \bigg[ T \bigg(C-\frac{\mathcal{R}}{S_0}\bigg) S_0^3 \bigg]_\F + \textrm{h.c.}  \bigg)\,,
\end{align}
such that $T$ acts as a Lagrange multiplier. Using the identity \eqref{eq:id1}, the Lagrangian \eqref{eq:starsugra2} can be brought to the form
\begin{align} \label{eq:starsugra:dual2}
    \mathcal{L} = - 3 \Big[ \Big( 1+T+\bar{T} - 6 \alpha  \, C\bar{C}  \Big) \, S_0 \bar{S}_0 \Big]_\D + 
    3 \Big( \big[C T \, S_0^3 \big]_\F + \textrm{h.c.}  \Big)\,,
\end{align}
which, after the field replacements
\begin{align} \label{eq:redef}
       C \rightarrow \frac{1}{\sqrt{6 \alpha}} \, C \,, \quad T \rightarrow T - \tfrac12
\end{align}
and the gauge--fixing \eqref{eq:gauge} takes the form of $\mathcal{N}=1$ supergravity coupled to a chiral superfield $T$ in the presence of a so-called stabiliser $C$, with a no--scale K\"ahler potential~\cite{Ellis:1984bm} and superpotential that are given by
\begin{align} \label{eq:starsukw}
  K  = -3 \ln \big(T + \bar{T} - C\bar{C} \big) \,,\quad W = M \, C \big(T-\tfrac12) \,,
\end{align}
respectively, where we have set $M:=\sqrt{ \tfrac{3}{2\alpha}}=3m\,$. 

Without going into details of our analysis below, inspection of \eqref{eq:starsukw} shows that $T=1/2$ and $C=0$ corresponds to a supersymmetric vacuum with vanishing superpotential and its first derivatives, describing the flat space vacuum in the geometric formulation. Note also that the theory has a global R-symmetry corresponding to phase transformations of $C$ under which the superpotential is charged and therefore $C=0$ corresponds to an extremum of the scalar potential, with $C$ being the goldstino superfield of spontaneously broken supersymmetry away from $T=1/2$.

Indeed, proceeding now to our analysis, the K\"ahler covariant derivatives of the superpotential read
\begin{align}
    \begin{pmatrix}
        D_T W \\ 
        D_C W
    \end{pmatrix} = \frac{M}{T+\bar{T}-C \bar{C}} \begin{pmatrix}
   C  \big(\bar{T}-2 T- C \bar{C}+\tfrac32\big) \\[0.5em]  \big( T-\tfrac12\big) \left(T+\bar{T}+2 C \bar{C}\right)   \end{pmatrix}
\end{align}
and the K\"ahler metric is given by
\begin{align}
\begin{pmatrix}
    K_{T \bar T} & K_{T \bar C} \\ K_{C \bar T} & K_{C \bar C}
\end{pmatrix}
=   \frac{3 }{\left(T+\bar{T}-C \bar{C}\right)^2}  \left(
\begin{array}{cc}
 1 & - C \\
 - \bar{C}& T+\bar{T} \\
\end{array}
\right)\,.
\end{align}
The scalar potential then takes the form
\begin{equation} \label{eq:potentialstar}
\begin{aligned}
V(T,\bar{T},C,\bar{C}) &=  \frac{M^2 }{12 \left(T+\bar{T}-C \bar{C}\right)^2} \Big\{1 -2(T+ \bar{T})+4 T \bar{T} + \big[8- 4  (T +  \bar{T}) \big] C \bar{C}\Big\}\,.
\end{aligned}
\end{equation}
Note from \eqref{eq:redef} that $T_{\mathtt R}>1/2$. For
\begin{align} \label{eq:bos}
    C=0 \,,\quad T_{\mathtt R} = \tfrac12 e^{\sqrt{\tfrac23} \phi}\,, \quad T_{\mathtt I} = 0 \,,
\end{align}
the Starobinsky scalar potential \eqref{eq:starpot} is recovered from the supergravity potential \eqref{eq:potentialstar}, so that the field $T$ plays the role of the complexified inflaton, with $M=3m$ fixing the scale of inflation. Moreover, at the point
\begin{align} \label{eq:min}
    \langle C \rangle=0\,,\quad \langle T \rangle = \tfrac12\,,
\end{align}
it holds that
\begin{align}
    D_C W = 0 = D_T W\,,
\end{align}
as well as 
\begin{align}
    \frac{\partial V}{\partial C} = 0 =  \frac{\partial V}{\partial T} \,,\quad
    \begin{pmatrix}
        \frac{\partial^2 V}{\partial T \partial \bar T} &   \frac{\partial^2 V}{\partial T \partial \bar C} \\[0.5em]
        \frac{\partial^2 V}{\partial C \partial \bar T} &   \frac{\partial^2 V}{\partial C \partial \bar C} 
    \end{pmatrix} = \tfrac13 M^2 \begin{pmatrix}
        1 & 0 \\ 0 & 1
    \end{pmatrix}\,, \quad V = 0 \,,
\end{align}
so that the expectation values \eqref{eq:min} correspond to a supersymmetric global Minkowski minimum of the potential \eqref{eq:potentialstar}, with the field $C$ playing the role of the goldstino (chiral) superfield, as we have already argued before.

On the other hand, during inflation, along the direction of $T_{\mathtt R}$ for $C={\bar C}=T_{\mathtt I}=0$, the masses--squared of the (canonically normalized) fields $C$ and $T_\mathtt I$ take the form
\begin{align}
    m_C^2 & = \frac{1}{K_{C\bar C}} \frac{\partial^2 V}{\partial C \partial \bar C} \Big|_{C=0=\bar C} = M^2\frac{ 1 + 2 (T+\bar T) - 2(T^2 +\bar T^2)}{18 \left(T+\bar{T}\right)^2} \\ 
    m_{T_\mathtt I}^2 &  = \frac{1}{K_{T_\mathtt I T_\mathtt I}} \frac{\partial^2 V}{\partial T_{\mathtt I}^2} \Big|_{C=\bar C=0=T_\mathtt I} = \tfrac29 M^2=8H^2\,,
\end{align}
where in the last equality, we used the relations $M=3m$ and $H=m/2$. Thus, $T_{\mathtt I}$ becomes heavy compared to the inflationary Hubble scale and decouples during inflation, while the inflaton mass vanishes  at large $T_{\mathtt R}$ as expected around a flat potential. We may thus set $T_{\mathtt I}=0$, implying that 
\begin{align}\label{mC2}
    m_C^2 & = \frac{M^2 (-4 T_{\mathtt R}^2 + 4T_{\mathtt R}+ 1)}{72 T_{\mathtt R}^2} \,.
\end{align}
It follows that the sgoldstino field $C$ becomes tachyonic for large $T_\mathtt R>(1+{\sqrt 2})/2$, therefore destabilising inflation. However, as it has been shown in~\cite{Antoniadis:2024ypf} and we shall demonstrate in the next subsections in the presence of an $R^3$ term and more generally of any additional higher power of the curvature scalar, the presence of the string dilaton renders $m_C^2$ positive, stabilising the inflaton potential at the R-symmetric point $C=0$.

Indeed, the dilaton is included by adding to the K\"ahler potential \eqref{eq:starsukw} the term $K_{\textrm{string}} = - \ln(S+\bar S)$ with no superpotential. The scalar potential is then modified by an additive contribution $|D_SW|^2=|W|^2$ and is multiplied by an overall factor $e^{K_\textrm{string}}=1/(S+\bar S)$. It follows that the expressions of $H^2$ and $m_{T_\mathtt I}^2$ are multiplied by the same factor so that $m_{T_\mathtt I}/H$ at $C=0$ is not modified by the presence of the dilaton. On the other hand, the additive contribution renders the (complex) sgoldstino non tachyonic with a mass~\cite{Antoniadis:2024ypf}:
\begin{align}\label{mC2dil}
m_C^2 & = \frac{M^2}{36(S+\bar S)T_{\mathtt R}^2}\left[ \left(T_{\mathtt R}-\frac12\right)^2 + 1\right]
\to\frac{M^2}{36(S+\bar S)}=H^2 \,,
\end{align}
where at the end of the line we considered the limit at large $T_{\mathtt R}$ corresponding to the inflation period. Thus the $C$ mass is equal to the Hubble scale, which makes its decoupling marginal and might have interesting imprints in non gaussianity, requiring a dedicated analysis. Of course, as we mentioned already, a dilaton stabilisation mechanism is needed to be implemented which is also necessary in order to study its possible effects during inflation.
 
%%%%%%%%%%%%%%%%%%%%%%%%%%%%%%%
\subsection{\texorpdfstring{Extending Starobinsky supergravity with an \ensuremath{R^3} term}{Extending Starobinsky supergravity with an R^3 term}} \label{subsec:mod}
%%%%%%%%%%%%%%%%%%%%%%%%%%%%%%%

To embed the ${R}^3$ term in $\mathcal{N}=1$ supergravity, we first note that, as in the case of $R^2$, it cannot be reproduced by an \F--density cubic in the chiral curvature superfield $\mathcal{R}$ and its conjugate, since the Ricci scalar $R$ appears in the upper component ($\Theta^2$ term) of $\mathcal{R}$. However, the superfield $ \mathscr{P}(\bar{\mathcal R})$, that is built as the chiral projector $\mathscr{P}$ acting on the conjugate of the curvature superfield $\bar{\mathcal{R}}$, has the Ricci scalar as its lowest component. Consequently, we consider the modification of \eqref{eq:starsugra} by the term $\mathcal{R}\bar{\mathcal{R}} \, \mathscr{P}(\bar{\mathcal R}) $, namely
\begin{align} \label{eq:modifsugra}
   \mathcal{L} = 3 \bigg[\bigg\{-1+ 6 \frac{\mathcal{R}}{S_0}\frac{\bar{\mathcal{R}}}{\bar{S}_0} \bigg[  \alpha - 3  \beta  \bigg(\mathscr{P}\bigg(\frac{\bar{\mathcal R}}{\bar{S}_0} \bigg)  + \bar{\mathscr{P}}   \bigg(\frac{  \mathcal R}{S_0}  \bigg)  \bigg) \bigg] \bigg\} S_0 \bar{S}_0  \bigg]_\D  \,,
\end{align}
whose bosonic part has precisely the $R$--dependence of the Lagrangian \eqref{eq:modif}. The chiral projector $\mathscr{P}$ (defined in the Appendix) consists of the supergravity generalisation of the ${\bar D}^2$ projector of global supersymmetry, which is the square of the (conjugate of) the super-covariant derivative that vanishes when acting on chiral superfields, so that $\mathscr{P}({\mathcal R})=0$ and $\mathscr{P}^2=0$.

We can then linearise \eqref{eq:modifsugra} by introducing, along with the chiral superfields $T$ and $C$ (see \eqref{eq:starsugra2}),  \textit{one} additional chiral superfield $X$ which we identify with $\mathscr{P}(\bar{\mathcal R})$, namely
\begin{equation}\label{eq:beystarsugra2}
\begin{aligned}
     \mathcal{L} = &  3 \Big[\big(-1+ 6 \alpha \, C\bar{C}   \big) \, S_0 \bar{S}_0  \Big]_\D \\
& \qquad +  3 \bigg( \bigg[ \bigg\{ T \bigg(C-\frac{\mathcal{R}}{S_0}\bigg) +  \tfrac{1}{2} \beta \, X \, C  \Big( X+12 \mathscr{P}(\bar{C} ) \Big)  \bigg\}\, S_0^3 \bigg]_\F + \textrm{h.c.}  \bigg)\,.
\end{aligned}
\end{equation}
Indeed, integrating out $T$ and $X$, the equation of motion of $X$ being
\begin{align} \label{eq:Xeom}
    X=-6 \mathscr{P}(\bar C)\,,
\end{align}
one can rewrite the second line of \eqref{eq:beystarsugra2} using
\begin{align}
-\big[CX^2\big]_\F + \textrm{h.c.}=6\big[\mathscr{P}(X C {\bar C})\big]_\F+\textrm{h.c.} =   6\big[(X+{\bar X}) C {\bar C} \big]_\D\,,
\end{align}
where in the last step we used the identity \eqref{eq:ids}. Thus, using $C={\cal R}/S_0$ and \eqref{eq:Xeom}, the Lagrangian \eqref{eq:beystarsugra2} is brought to the form \eqref{eq:modifsugra}. Note that the lower component of $X$ is the scalar curvature $R$. On the other hand, if we apply the identity \eqref{eq:id1} without integrating out $T$ and $X$, we find the dual form:
\begin{equation}\label{eq:beystarsugraintout}
\begin{aligned}
     \mathcal{L} & =   - 3 \Big[\Big\{1 + T +\bar{T} - 6 C\bar{C} \big[\alpha + \beta  ( X +\bar{X}) \big] \Big\} \, S_0 \bar{S}_0  \Big]_\D   \\
     & \qquad + 3  \bigg( \Big[  C \Big(  T + \tfrac12 \, \beta X^2  \Big) \, S_0^3 \Big]_\F + \textrm{h.c.}  \bigg)\,,
\end{aligned}
\end{equation}
which, after the redefinition  \eqref{eq:redef}, leads to $\mathcal{N}=1$ supergravity with K\"ahler potential and superpotential given by
\begin{align} \label{eq:kaehler1}
  K  &= -3 \ln \Big\{T + \bar{T} - C\bar{C} \Big[1+ \tfrac{1}{m^2}  \beta  ( X +\bar{X}) \Big]  \Big\} \\ \label{eq:superpot1}
  W & = 3 m \, C \Big[  \big(T-\tfrac12) + \tfrac{1}{8m^4}  \beta X^2 \Big]  \,
\end{align}
respectively, where we have used again the rescaling $\beta \rightarrow \tfrac{1}{9m^4} \beta$ such that $\beta$ is now dimensionless and we have also rescaled the auxiliary field as $X\to\frac{3}{2}X$. Note that $X$ has no quadratic kinetic term around $C=0$ and can be treated as auxiliary field that should be integrated out~\cite{Cecotti:1987sa}. Indeed, only around the vacuum $C=0$ the supergravity action contains the gravitational $R^3$ theory. Moreover, positivity of $C$ kinetic terms imply:
\begin{align} \label{positivityC}
-\beta ( X +\bar{X})<m^2\,.
\end{align}

Indeed, the K\"ahler derivatives read
\begin{align} \label{eq:kaehlerder}
        \begin{pmatrix}
        D_T W \\ 
        D_C W \\
        D_X W
    \end{pmatrix} &= \frac{1}{ 8\big[ m^2 \left(T+\bar{T} -C \bar{C}\right)-  \beta \, C \bar{C} \left(X+\bar{X}\right) \big] } \\
 & \nonumber \quad  \times  \begin{pmatrix}
- 3 \, C \, \Big[ 4 m^3 \left(2 C \bar{C}-2 \bar{T}+4 T-3\right)+8 \beta  m\, C \bar{C} \left(X+\bar{X}\right)+\frac{3\beta}{m}  X^2 \Big] \\[0.6em] 3 \Big[ 8 m\, \big(T-\tfrac12 \big)+\tfrac{\beta}{ m^3}  X^2 \Big] \Big[ m^2 (T+\bar{T} + 2 C \bar{C}) + 2\beta  \,C \bar{C} \left(X+\bar{X}\right) \big] \\[0.6em]   3 \beta \, C\, \Big\{  \tfrac{2}{m} X \left(T + \bar{T}  \right)+C \bar{C} \Big[ 24 m \big( T-\tfrac12 \big) - \tfrac{2}{m} X+ \tfrac{\beta}{m^3}  X \left(X-2 \bar{X}\right) \Big] \Big\}
  \end{pmatrix}
\end{align}
and the K\"ahler metric is given by
\begin{align} \label{eq:kaehlerR3}
\begin{pmatrix}
    K_{T \bar T} & K_{T \bar C} & K_{T \bar X}  \\ K_{C \bar T} & K_{C \bar C} & K_{C \bar X}  \\ K_{X \bar T} & K_{X \bar C} & K_{X \bar X} 
\end{pmatrix}
& = \frac{3m^2}{\big[ m^2 \left(T+\bar{T} -C \bar{C}\right)-  \beta \, C \bar{C} \left(X+\bar{X}\right) \big]^2}\\[0.5em]
 & \hspace{-3.5cm} \nonumber  \times \begin{pmatrix} \displaystyle
     m^2 &   - C \big[  m^2 +   \beta  \left(X+\bar X\right)\big] &  - \beta \, C \bar{C} \\[0.5em]
 - \bar{C}\big[  m^2 +   \beta  \left(X+\bar X\right)\big] &    \left( T + \bar T \right) \big[  m^2 +  \beta  \left(X+\bar X\right)\big] &  \beta   \,\bar{C} \left(T+ \bar{T} \right) \\[0.5em]
-  \beta   \, C \bar{C} &  \beta \, C \left(T+ \bar T \right) &  \big(\tfrac{\beta}{m}\big)^2 (C \bar{C})^2 
\end{pmatrix}\,.
\end{align}
The corresponding scalar potential then takes the form
\begin{align}\label{potentialR3}
\begin{aligned}
V(T,C,X,\bar{T},\bar{C},\bar X) &= \frac{3}{32 \big[  m^2 \left(T+\bar{T} -C \bar{C}\right)-\beta \, C \bar{C} \left(X+\bar{X}\right) \big]^2} \\
& \quad \times \bigg\{4 m^4 (2X \bar{T}+ 2T \bar{X} - X-\bar{X}) - 2m^2 X \bar{X}   - \beta \, X\bar{X} ( X+ \bar{X})     \\
& \qquad \qquad - 8  m^2 \,  C \bar{C}\Big[ 4m^4(T+ \bar{T}- 2 ) + \beta   (X^2 + \bar{X}^2)\Big]\bigg\}\,.
\end{aligned}
\end{align}

Splitting the three superfields $T$, $C$ and $X$ into their real and imaginary parts, the K\"ahler metric $\mathcal{K}$ of the corresponding 6 real degrees of freedom (d.\ o.\ f.) takes the form
\begin{align} \label{eq:kaehlerR3real}
\begin{pmatrix}
    \mathcal{K}_{T_{\mathtt R}  T_{\mathtt R}} &  \mathcal{K}_{T_{\mathtt R}  C_{\mathtt R}}  &  \mathcal{K}_{T_{\mathtt R}  X_{\mathtt R}} &   \mathcal{K}_{T_{\mathtt R} T_{\mathtt I}} &  \mathcal{K}_{T_{\mathtt R}  C_{\mathtt I}}  &  \mathcal{K}_{T_{\mathtt R}  X_{\mathtt I}} 
    \\ \mathcal{K}_{C_{\mathtt R} T_{\mathtt R}} &  \mathcal{K}_{C_{\mathtt R}  C_{\mathtt R}}  &  \mathcal{K}_{C_{\mathtt R}  X_{\mathtt R}} &   \mathcal{K}_{C_{\mathtt R}  T_{\mathtt I}} &  \mathcal{K}_{C_{\mathtt R}  C_{\mathtt I}}  &  \mathcal{K}_{C_{\mathtt R}  X_{\mathtt I}} 
    \\ \mathcal{K}_{X_{\mathtt R}  T_{\mathtt R}} &  \mathcal{K}_{X_{\mathtt R}  C_{\mathtt R}}  &  \mathcal{K}_{X_{\mathtt R}  X_{\mathtt R}} &   \mathcal{K}_{X_{\mathtt R}  T_{\mathtt I}} &  \mathcal{K}_{X_{\mathtt R}  C_{\mathtt I}}  &  \mathcal{K}_{X_{\mathtt R} X_{\mathtt I}} 
 \\   \mathcal{K}_{T_{\mathtt I}  T_{\mathtt R}} &  \mathcal{K}_{T_{\mathtt I} C_{\mathtt R}}  &  \mathcal{K}_{T_{\mathtt I}  X_{\mathtt R}} &   \mathcal{K}_{T_{\mathtt I} T_{\mathtt I}} &  \mathcal{K}_{T_{\mathtt I}  C_{\mathtt I}}  &  \mathcal{K}_{T_{\mathtt I}  X_{\mathtt I}} 
  \\ \mathcal{K}_{C_{\mathtt I} T_{\mathtt R}} &  \mathcal{K}_{C_{\mathtt I}  C_{\mathtt R}}  &  \mathcal{K}_{C_{\mathtt I}  X_{\mathtt R}} &   \mathcal{K}_{C_{\mathtt I}  T_{\mathtt I}} &  \mathcal{K}_{C_{\mathtt I}  C_{\mathtt I}}  &  \mathcal{K}_{C_{\mathtt I}  X_{\mathtt I}} 
  \\ \mathcal{K}_{X_{\mathtt I}  T_{\mathtt R}} &  \mathcal{K}_{X_{\mathtt I} C_{\mathtt R}}  &  \mathcal{K}_{X_{\mathtt I} X_{\mathtt R}} &   \mathcal{K}_{X_{\mathtt I}  T_{\mathtt I}} &  \mathcal{K}_{X_{\mathtt I}  C_{\mathtt I}}  &  \mathcal{K}_{X_{\mathtt I}  X_{\mathtt I}} 
  \end{pmatrix}
& =  \begin{pmatrix} \mathbf{K}_{3\times 3}  & \mathbf{0}_{3\times 3} \\
 \mathbf{0}_{3\times 3} & \mathbf{K}_{3\times 3} 
\end{pmatrix}\,,
\end{align}
where $\mathbf K_{3\times 3} $ is the matrix \eqref{eq:kaehlerR3} written in terms of the $6$ real component fields of $T$, $C$ and $X$. Consequently, the determinant of the matrix $\mathcal K$ is given by
\begin{align}
    \det \mathcal K = (\det \mathbf K)^2 = \frac{729 \,\beta ^4 m^8 \,\big( C_{\mathtt R}^2 +C_{\mathtt I}^2\big)^2}{\big[  m^2 \left(2T_{\mathtt R} - C_{\mathtt R}^2 - C_{\mathtt I}^2\right) - 2 \beta \, X_{\mathtt R}  ( C_{\mathtt R}^2 +C_{\mathtt I}^2) \big]^8}\,,
\end{align}
so that the metric $\mathcal K$ is degenerate to $0$-th order in $C \bar C$, which means that at least one field direction has no kinetic term to this order; actually this is the complex auxiliary field $X$ around $C=0$. 

\begin{figure}[htbp]
\centering
\includegraphics[scale=0.9]{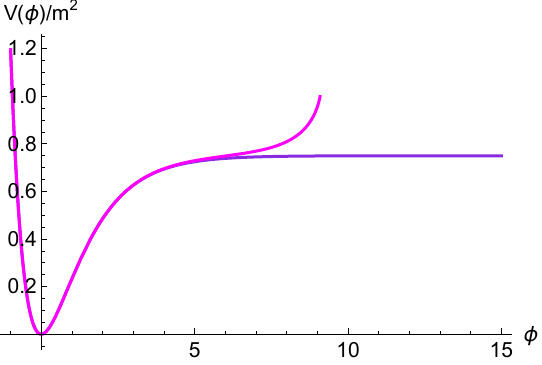}
\caption{The inflaton potential (over $m^2$) of $ R^2$ supergravity (in \textcolor{brightpurple}{purple}) versus that of $ R^2 +   R^3$ supergravity (in \textcolor{brightmagenta}{magenta}) for $\beta= -5 \times 10^{-5}$.} \label{fig:potR3}
\end{figure}

Remaining at the same order, the equation of motion (e.\ o.\ m.) of $X$, $\partial_X V=0$, from \eqref{potentialR3} yields the equation:\footnote{This is equivalent with integrating out $X$, since as $C\to 0$ the kinetic term of $X$ vanishes and thus its mass diverges.}
\begin{align} \label{eq:eomX}
-4m^4(2{\bar T}-1)+{\bar X}(2m^2+\beta{\bar X})=-2\beta X{\bar X}\,,
\end{align}
that brings the potential at $0$-th order in $C \bar C$ in the form
\begin{align} \label{potential0}
V(C=0)=\frac{3X{\bar X}}{16m^4(T+{\bar T})^2}[m^2+\beta(X+{\bar X})]\,,
\end{align}
which is positive definite due to the constraint \eqref{positivityC}. The real and imaginary part of \eqref{eq:eomX} lead to the equations:
\begin{align} \label{eq:eom1}
-8 m^4 \big( T_{\mathtt R} -\tfrac12 \big) +2 m^2 X_{\mathtt R}+ \beta  \left(X_{\mathtt I}^2+3 X_{\mathtt R}^2\right) &=0 \\ \label{eq:eom2}
  -  4 m^4 \, T_{\mathtt I} +  m^2 \,X_{\mathtt I} + \beta\, X_{\mathtt I} X_{\mathtt R} & = 0 \,,
\end{align}
which express the auxiliary field $X$ in terms of the propagating d.\ o.\ f. $T$ and $C$. The second equation implies that $X_I$ is proportional to $T_I$ and using it in \eqref{eq:eomX} we find the mass of the field $T_{\mathtt I}$ to be (to $0$-th order in $C$)
\begin{align}
    m_{T_{\mathtt I}}^ 2 = 2\frac{m^4( m^2+  2\beta X_{\mathtt R})}{\left( m^2+  \beta X_{\mathtt R} \right)^2}\,,
     \end{align}
taking into account the normalisation of its kinetic terms, which is positive. 
Since we are interested in small reformations of the Starobinsky model (see $\beta\lesssim{\cal O}({10^{-4}})$ in Fig.~\ref{fig:potR3}), the relations between the masses and $H$ remain essentially the same.
Thus, the field  $T_{\mathtt I}$ is heavier than $H$ during inflation and can be set to zero,
\begin{align} \label{eq:xI}
 T_{\mathtt I} = 0=X_{\mathtt I}\,.
\end{align}
The e.\ o.\ m. \eqref{eq:eom1} then takes the form of second--order equation
\begin{align} 
  3\beta X_{\mathtt R}^2 +2 m^2  X_{\mathtt R} - 8 m^4 \big(T_{\mathtt R} -\tfrac12\big) =0 \,,
\end{align}
which has the two solutions
\begin{align} \label{eq:xR}
   X_{\mathtt R}^\pm=   -\tfrac{ m^2}{3 \beta } \Big[ 1\pm\sqrt{1+24 \beta \, \big(T_{\mathtt R} -\tfrac12\big)} \,\Big] \,,
\end{align}
where we must impose \eqref{positivityC} and that the argument of the square root is non--negative. Obviously \eqref{positivityC} is automatically satisfied for $\beta<0$ while for $\beta$ positive it is easy to see that only $X_{\mathtt R}^-$ satisfies the conditions. On the other hand, the solution $X_{\mathtt R}^+$ is excluded since it has no limit at $\beta\to 0$.

With the fields $T_{\mathtt I}=X_{\mathtt I}=0$ and $X_{\mathtt R}$ given by \eqref{eq:xR} the scaler potential \eqref{potentialR3} takes the form
\begin{align}
\begin{aligned}\label{potentialR3eff}
V_{\textrm{eff}}(T_{\mathtt R}, C, \bar C) &=   m^2\, V_0(T_{\mathtt R}) +  m^2  \, V_1 (T_{\mathtt R})\, C\bar C + \mathcal{O}\big((C \bar C)^2 \big) \,,
\end{aligned}
\end{align}
evaluated at $X=X_{\mathtt R}^-$, where
\begin{align} \label{eq:v0}
V_0 (T_{\mathtt R}) &= \frac{1}{576 \beta^2\,T_{\mathtt R}^2} \left(\sqrt{24 \beta \big( T_{\mathtt R} -\tfrac12 \big)+1} - 1\right)^2 \left( \,2 \sqrt{24 \beta \big( T_{\mathtt R} -\tfrac12 \big) +1}+1\right)
 \end{align}
 and
 \begin{align}
\begin{aligned}
    V_1(T_{\mathtt R}) & = \frac{1}{864 \, T_{\mathtt R}^3} \bigg\{ \frac{1}{\beta^2}\Big[ 1- \sqrt{1+ 24 \beta \big( T_{\mathtt R} -\tfrac12 \big) } \,\Big]\\
    & \quad -\frac{6}{ \beta} \Big[ 5 +2  T_{\mathtt R} - 4 \big( T_{\mathtt R} +1 \big) \sqrt{1+ 24 \beta \, \big( T_{\mathtt R} -\tfrac12 \big) } 
 + 24 \beta \, \big( 7T_{\mathtt R}^2  -4 T_{\mathtt R} -2 \big)  \Big] \bigg\}  \,.
\end{aligned}
   \end{align}
Note that the expression \eqref{eq:v0} of $V_0$ for $T_{\mathtt R}=\frac{1}{2}e^{\sqrt{2/3}\phi}$ coincides with the Starobisnky potential deformed by the addition of an $R^3$ term in \eqref{eq:vstarmod}. 
For small $|\beta|$, we have  
\begin{align}
\begin{aligned} \label{eq:v0smallbeta}
V_0(T_{\mathtt R}) &= \frac{3  \big( T_{\mathtt R}-\tfrac12 \big)^2}{4 \,T_{\mathtt R}^2} - \frac{3 \big( T_{\mathtt R}-\tfrac12 \big)^3}{ T_{\mathtt R}^2}\,  \beta+ \mathcal{O} \left(\beta^2\right)\\[0.5em]
V_1(T_{\mathtt R}) &=  \frac{3(-4T_{\mathtt R}^2+ 4T_{\mathtt R}+1)}{16\, T_{\mathtt R}^3}- \frac{3\big( T_{\mathtt R}-\tfrac12 \big)^2 \big(  T_{\mathtt R} +  \tfrac12 \big) }{T_{\mathtt R}^3} \, \beta  + \mathcal{O} \left(\beta^2\right)\,.
\end{aligned}
 \end{align}

The point \eqref{eq:min} corresponds to a supersymmetric global Minkowski minimum like in Starobinsky supergravity. Indeed
$X_{\mathtt R}^-$ vanishes due to \eqref{eq:xR} as $X_{\mathtt R}^-\simeq 4m^2(T_{\mathtt R}-1/2)$ and the potential \eqref{potentialR3eff} becomes (see \eqref{potentialR3})
\begin{align}
V_{\textrm{eff}}\simeq \frac{3m^2}{4T_{\mathtt R}^2}\left[\left(T_{\mathtt R}-\tfrac{1}{2}\right)^2+\frac{1}{2T_{\mathtt R}} C{\bar C}\right]\,,
\end{align}
while it also holds that
\begin{align}
     D_T W = D_C W  =  0  \,.
\end{align}
Thus \eqref{eq:min} is a supersymmetric minimum at zero energy with
\begin{align}
     \frac{\partial^2 V_{\textrm{eff}}}{\partial T_{\mathtt R} \partial T_{\mathtt R} } = 6 m^2 >0 \,, \quad
    \frac{\partial^2 V_{\textrm{eff}}}{\partial C \partial \bar C} = 3 m^2 >0\,
\end{align}
corresponding to physical masses $m_T=m_C=m$.
However, away from $T_{\mathtt R}=1/2$, $C$ becomes tachyonic, as it was already the case in \eqref{mC2} for vanishing $\beta$.
To linear order in $|\beta|<1$, the mass--squared of $C$ reads
\begin{align}
\begin{aligned}
   m_C^2 &= \frac{1}{K_{C\bar C}}   \frac{\partial^2 V_{\textrm{eff}}}{\partial C \partial \bar C} \Big|_{C=0=\bar C} \\
   & =\frac{m^2}{8} \left(\frac{1}{T_{\mathtt R}^2}+\frac{4}{T_{\mathtt R}}-4\right) + \frac{  m^2 \left(8 T_{\mathtt R}^3-20 T_{\mathtt R}^2+6 T_{\mathtt R}+1\right)}{4\, T_{\mathtt R}^2}\,\beta + \mathcal{O}\left(\beta^2\right) \,.
\end{aligned}
\end{align}

In \cite{Antoniadis:2024ypf}, the problem of the tachyonic mass of $C$ arising from the form of the scalar potential \eqref{eq:potentialstar} in the simplest model of Starobinsky supergravity \eqref{eq:starsugra} was cured by introducing the string dilaton $\Phi$ via its chiral superfield $S$ with $S_{\mathtt R}=e^{-2\Phi}$. In our case, this amounts to adding to the K\"ahler potential \eqref{eq:kaehler1} a term that reads
\begin{align}
    K_{\textrm{string} } = - \ln(S+\bar S)\,,
\end{align}
while the superpotential \eqref{eq:superpot1} remains unchanged. The K\"ahler derivatives \eqref{eq:kaehlerder} then remain unchanged and we additionally have
\begin{align}
    D_S W = -\frac{3C  \big[8 m^4 \big( T-\tfrac12 \big) +\beta  X^2 \big] }{8  m^3 \left(S+\bar{S}\right)}\,,
\end{align}
while the K\"ahler metric now takes the form
\begin{align} \label{eq:kaehlerR3dil}
\begin{pmatrix}
    K_{T \bar T} & K_{T \bar C} & K_{T \bar X}  & K_{T \bar S}   \\ K_{C \bar T} & K_{C \bar C} & K_{C \bar X}   & K_{C \bar S}  \\ K_{X \bar T} & K_{X \bar C} & K_{X \bar X}  & K_{X \bar S}  \\ K_{S \bar T} & K_{S \bar C} & K_{S \bar X}  & K_{S \bar S}  
\end{pmatrix}
& = \begin{pmatrix}
\multicolumn{3}{c}{\multirow{3}{*}{$\mathbf{K}_{3\times 3}$}} & 0 \\
 & & & 0 \\
 & & & 0 \\
0 & 0 & 0 & \dfrac{1}{(S+\bar S)^2}
\end{pmatrix} \,,
\end{align}
where $\mathbf{K}_{3\times 3}$ is the K\"ahler metric \eqref{eq:kaehlerR3} in the absence of the dilaton. The new scalar potential $V(T,C,X,S,\bar{T},\bar{C},\bar X, \bar S)$ then reads
\begin{align}
\begin{aligned}
      V &=  \frac{1}{64 \left(S+\bar{S}\right) \Big\{- m^2 \left(T+\bar{T}\right) +C \bar{C} \big[  m^2 + \beta  \left(X+\bar{X}\right)\big]\Big\}^3}  \\
& \quad \times \bigg\{ 6 m^2 \left(T+\bar{T}\right) \Big[-4 m^4 \left(2 X \bar{T}+2 T \bar{X}-X-\bar{X}\right)+ X \bar{X}\big[ 2 m^2 +\beta  \left(X+\bar{X}\right)\big] \Big]\\
 &  \qquad \,\,  +3 \, C \bar{C} \Big[16 m^8 \left(-4 T \bar{T}+4 \bar{T}^2-2 \bar{T}+4 T^2-2 T-3\right) - 8 m^6 (-2 X \bar{T}\\
& \qquad  \,\,  -2 T \bar{X} +X+\bar{X}) +  4 m^4 \big[ \beta  X^2 \left(2 \bar{T}+4 T+1\right)+X \bar{X} \big(4 \beta  (T+\bar{T}-1)  - 1 \big) \\
& \qquad \,\, +  \beta   \bar{X}^2 \left(4 \bar{T}+2 T+1\right)\big] - \beta  X \bar{X} \Big( 6m^2 \left(X+\bar{X}\right)+\beta   (7 X \bar{X}+2 X^2+2 \bar{X}^2 ) \Big) \Big]  \\
&  \qquad \,\,  - 48 m^2 \, (C \bar{C})^2 \Big[ m^2+\beta  \left(X+\bar{X}\right)\Big] \Big[ 4 m^4 (T+\bar{T}-2) +\beta  \left(X^2+\bar{X}^2\right)\Big] \bigg\}\,.
\end{aligned}
\end{align}

To $0$-th order in $C \bar C$, the e.~o.~m.~of the real fields $X_{\mathtt R}$ and $X_{\mathtt I}$ are precisely the same as without the dilaton, namely \eqref{eq:eomX}, or equivalently \eqref{eq:eom1} and \eqref{eq:eom2}; the same is also true for the solution $X_{\mathtt R}^-$, given in \eqref{eq:xR}. The effective potential is now given by
\begin{align} \label{eq:veffdil}
V_{\textrm{eff}}(T_{\mathtt R}, C, S,\bar C,\bar S) &=   m^2\, V_0(T_{\mathtt R},S,\bar S) +  m^2  \, V_1 (T_{\mathtt R}, S, \bar S)\, C\bar C + \mathcal{O}\big((C \bar C)^2 \big) \,,
\end{align}
where
\begin{align}
     V_0(T_{\mathtt R},S,\bar S) = \frac{1}{S+\bar S}  V_0(T_{\mathtt R})\,,
\end{align}
with $ V_0(T_{\mathtt R})$ given in \eqref{potential0} or \eqref{eq:v0} and
\begin{align}
V_1 (T_{\mathtt R}, S, \bar S) =\frac{12\Big[4m^4 T_{\mathtt R} -16X_{\mathtt R}^- \big(m^2+2\beta X_{\mathtt R}^-\big)\Big]^2 
+ 3(X_{\mathtt R}^-)^2 \Big(m^2+2\beta X_{\mathtt R}^-\Big)^2}{128 m^8 \, T_{\mathtt R}^3 \left(S+\bar{S}\right)}  
\,,
\end{align}
with $X_{\mathtt R}^-$ given in \eqref{eq:xR}. Thus, the mass-squared of the sgoldstino $C$
is positive definite and the tachyonic instability is \textit{cured}.\footnote{Of course, an appropriate stabilisation mechanism of the dilaton is required~\cite{Antoniadis:2024ypf}.} 
Its mass being of order $H$ requires a dedicated analysis of its effects, as we already mentioned in section~\ref{subsec:Starsugra}.

As discussed in Section~\ref{sec:bosonic}, the physically interesting case is for $\beta<0$, where the potential has a sharp rising forbidding large inflaton excursions. In this case, positivity of the argument of the square root in \eqref{eq:xR} implies an upper bound on $T_{\mathtt R}$ which, upon the field redefinition \eqref{eq:bos}, takes the form
\begin{align}
    \phi \leq \sqrt{\tfrac32} \ln\left(1-\frac{1}{12\beta}\right)\,.
\end{align}
This bound is a consequence of the condition $F''>0$ of \eqref{eq:conditionsf} and is identical to the bound of the bosonic case \cite{Gialamas:2025ofz} as should be. As mentioned already above, the expression $m^2\, V_0(T_{\mathtt{R}})$ matches precisely the potential \eqref{eq:vstarmod} of $\alpha R^2  +\beta R^3$ gravity, %so the latter is precisely reproduced by the $0$-th--order term in $C \bar C$ of the effective potential \eqref{eq:veffdil}, 
modulo the dilaton prefactor. In Figure \ref{fig:potR3}, we plot $  V_0\left( T_{\mathtt R}(\phi) \right)$ and compare it with the Starobinsky potential $ V_{\textrm{Star}}(\phi)/m^2$ (obtained for $\beta=0$) given in \eqref{eq:starpot}. 
A small negative value of $\beta$ rises the potential before $\phi$ reaches the critical value $\phi_c\simeq 10$ at which the mass of the KK tower predicted by the Swampland distance conjecture becomes of order the inflation scale and violates unitarity~\cite{Antoniadis:2024ypf} (see discussion in Section~\ref{sec:bosonic}).
One sees that $|\beta|\lesssim{\cal O}(10^{-4})$ is needed, which is also compatible with ACT observations, leading to the question of how accidental inflation was~\cite{Antoniadis:2025pfa}. Indeed, identifying $m$ with the scale of validity of the inflation EFT, $\beta$ is a dimensionless parameter that deforms slightly the Starobinsky model for both theoretical and experimental reasons. The origin of its smallness remains an open question.

%%%%%%%%%%%%%%%%%%%%%%%%%%%%%%%
\subsection{\texorpdfstring{The \ensuremath{f(R)} supergravity}{The f(R) supergravity}} \label{subsec:f(R)}
%%%%%%%%%%%%%%%%%%%%%%%%%%%%%%%

As we saw previously, the supersymmetric embedding of $R+R^2$ gravity requires a $\D$--density involving a \textit{real} function $\mathcal N$ such that \cite{Cecotti:1987sa,Ferrara:2013wka,Ketov:2013sfa}
\begin{align}
    \frac{\partial^2 \mathcal N}{\partial \mathcal R \partial{\bar{ \mathcal R}}} \neq 0\,.
\end{align}
Turning now to the generalisation of arbitrarily many higher powers of the Ricci scalar $R$ (without extra derivatives), it is again a real function that may be employed to construct the  supergravity embedding of $F(R)$ gravity \cite{Cecotti:1986qw,Cecotti:1987sa}. 
We will now show that the generalisation of the $R^3$ construction \eqref{eq:modifsugra} is straightforward using the chiral projector 
$-6\mathscr{P}(\bar{\mathcal R})$, whose lower component is the scalar curvature $R$. Introducing an analytic function $f(X)$ as
\begin{align} \label{eq:deff}
f(X):=\sum_{n=1}^\infty \alpha_nX^n := Xh(X)\,,
\end{align}
such that $\alpha_1\equiv\alpha$ and $\alpha_2\equiv\beta$, where $h(X)$ is also an analytic function of $X$,  one can write the Lagrangian
\begin{equation}\label{eq:f(R)sugra4}
\begin{aligned}
     \mathcal{L} = & - 3 \big[S_0 \bar{S}_0  \big]_\D  - \tfrac{3}{2} \bigg( \bigg[\mathcal R  \, f \bigg(\!-6\mathscr{P} \bigg(\frac{\bar{\mathcal R}}{\bar{S}_0} \bigg) \bigg) \, S_0^2 \bigg]_\F + \textrm{h.c.}  \bigg)\,,
\end{aligned}
\end{equation}
whose bosonic component is 
\begin{align}\label{eq:f(R)sugra4bos}
     \mathcal{L}_\text{bos} =\tfrac12 \sqrt{-g} \big[R+Rf(R)\big] \,.
\end{align}
Comparing the Lagrangians \eqref{eq:f(R)sugra4bos} and  \eqref{eq:fR}, we find that the relation between the function of $F(R)$ gravity and our function $f$ reads
\begin{align}
\label{eq:ftoF}
    F(R) =R + R f(R)  \,.
\end{align}
Using the definition \eqref{eq:deff} and the identity \eqref{eq:ids}, one can now rewrite the Lagrangian \eqref{eq:f(R)sugra4} as a $\D$-density
\begin{align} \label{eq:f(R)sugra2}
        \mathcal{L} = 3  \bigg[\bigg\{-1+ 3\frac{\mathcal{R}}{S_0}\frac{\bar{\mathcal{R}}}{\bar{S}_0} \bigg[ h\bigg(\!-6\mathscr{P} \bigg(\frac{\bar{\mathcal R}}{{\bar S}_0} \bigg) \bigg)+ \bar{h}\bigg( -6\bar{\mathscr{P}}\bigg(\frac{\mathcal R}{S_0}\bigg) \bigg) \bigg] \bigg\} S_0 \bar{S}_0  \bigg]_\D  \,.
\end{align}

We next linearise the above Lagrangian by introducing \textit{three} chiral superfields $T,C,X$ like in the $R^3$ case, so that we can rewrite \eqref{eq:f(R)sugra2} in the dual form
\begin{equation}\label{eq:f(R)sugra3}
\begin{aligned}
     \mathcal{L} = & - 3 \big[S_0 \bar{S}_0  \big]_\D \\
& \quad + 3 \bigg( \bigg[ \bigg\{ T \bigg(C-\frac{\mathcal{R}}{S_0}\bigg) -\tfrac12  C \Big(\!-6f'(X) \mathscr{P}(\bar C) +f(X) -X f'(X)\Big)  \bigg\}\, S_0^3 \bigg]_\F + \textrm{h.c.}  \bigg)\,.
\end{aligned}
\end{equation}
Indeed, the e.\ o.\ m.\ of the superfield $X$ takes the form
\begin{align} \label{eq:Xeomf}
X=-6\mathscr{P}(\bar C)=-6\mathscr{P}\bigg(\frac{\bar{\mathcal{R}}}{\bar{S}_0} \bigg)
\end{align}
like \eqref{eq:Xeom} in $R^3$ supergravity, under the condition that $f''(X) \neq 0$. When the function $f(X)$ is linear, $f(X)=\alpha X$, the superfield $X$ drops from the expression \eqref{eq:f(R)sugra3}, which reproduces the $R^2$ supergravity \eqref{eq:starsugra} and \eqref{eq:starsugra2}, fixing the 
normalisation of the terms in the parenthesis multiplying $C$. Setting now $C=\mathcal{R}/S_0$ as imposed by the Lagrange multiplier $T$, and using \eqref{eq:Xeomf}, the action \eqref{eq:f(R)sugra3} reduces to \eqref{eq:f(R)sugra2} or equivalently \eqref{eq:f(R)sugra4}.

On the other hand, applying the identity \eqref{eq:ids} without integrating out the fields $T$ and $X$, we find that the Lagrangian \eqref{eq:f(R)sugra3} corresponds to matter--coupled $\mathcal{N}=1$ supergravity with  K\"ahler potential and superpotential given by 
\begin{align} 
\label{eq:kf}
    K &= -3 \ln \Big\{ T +\bar{T} -C \bar{C} \big[f'(X) + \bar{f}'(\bar{X})\big] \Big\} \\ \label{eq:wf}
    W &= \sqrt{3} C \Big\{ T-\tfrac{1}{2} -\tfrac12 \big[f(X)- X f'(X)\big] \Big\}
\end{align}
respectively, after redefining $T \rightarrow T -\tfrac12$ and $C\to C/\sqrt{3}$. It follows that the global R-symmetry of Starobinsky supergravity is present for a general function $f(R)$, since the superpotential is still linear in $C$.
Note that positivity of the metric of $C{\bar C}$ implies
\begin{align}\label{positivityCgen}
  f_\mathtt{R}'(X) >0\,,
\end{align}
while, as in the case of $R^3$, $X$ is non-propagating auxiliary field around $C=0$ and should be expressed as a function of the propagating d.\ o.\ f.\ of the superfields $T$ and $C$. Indeed, the K\"ahler derivatives read
\begin{align} \label{eq:kaehlerderG}
        \begin{pmatrix}
        D_T W \\ 
        D_C W \\
        D_X W
    \end{pmatrix} &= \frac{\sqrt{3}}{2 \Big[ T+\bar{T} -C \bar{C} \left(f'(X) + \bar{f}'(\bar{X})\right)\Big]} \\
 & \nonumber \,  \times  \begin{pmatrix}
 C\Big[ 3  -4 T +2 \bar{T} + 3 f(X) -3 X f'(X) - 2 C \bar{C} \big( f'(X)+\bar{f}'(\bar{X}) \big) \Big]  \\[0.6em]  \Big[ 2T - 1 - f(X) + X f'(X)  \Big] \Big[ T +\bar{T} +2 C \bar{C} \left(f'(X)+ \bar{f}'(\bar{X})\right) \Big] \\[0.6em]   C f''(X) \Big\{  X (T+\bar{T}) +  C  \bar{C} \Big[ 6T-3 \ -  3 f(X) - X \big(- 2 f'(X) + \bar{f}'(\bar{X})\big)  \Big]  \Big\}
  \end{pmatrix}
\end{align}
and the K\"ahler metric is given by
\begin{align} \label{eq:kaehlerG}
\begin{pmatrix}
    K_{T \bar T} & K_{T \bar C} & K_{T \bar X}  \\ K_{C \bar T} & K_{C \bar C} & K_{C \bar X}  \\ K_{X \bar T} & K_{X \bar C} & K_{X \bar X} 
\end{pmatrix}
& = \frac{3}{\Big[ T+\bar{T} -C \bar{C}\left(f'(X) + \bar{f}'(\bar{X})\right) \Big]^2}\\[0.5em]
 & \hspace{-0.5cm} \nonumber  \times \begin{pmatrix} \displaystyle
    1 &    - C \left(f'(X) + \bar{f}'(\bar{X})\right)   &   - C \bar{C} \bar{f}''(\bar X) \\[0.5em]
   -\bar{C} \left(f'(X) + \bar{f}'(\bar{X})\right)  &   \left( T + \bar T \right)  \left(f'(X) + \bar{f}'(\bar{X})\right)  & \bar{C} \left(T+ \bar{T} \right) \bar{f}''(\bar X) \\[0.5em]
   -C \bar{C} f''(X) & C \left(T+ \bar T \right) f''(X) & (C \bar{C})^2 f''(X) \bar{f}''(\bar X)
\end{pmatrix}\,.
\end{align}
We can then compute the scalar potential to find
\begin{align}\label{potf}
        V &=   \nonumber \frac{1}{4 \bigg\{T+\bar{T} -C \bar{C}\, \Big[f'(X) + \bar{f}'(\bar{X})\Big]\bigg\}^2} \times \bigg\{ -(X+\bar{X})+2 \big(X \bar{T}+ T \bar{X}\big) - \Big[ X \bar{f} (\bar{X})+ \bar{X} f(X) \Big]  \\
&\qquad  \qquad +4 C \bar{C}\, \Big[2 - (T+\bar{T})+ f(X) +\bar{f}(\bar{X}) -\big[ X f'(X) +  \bar{X} \bar{f}'(\bar{X})\big] \Big]   \bigg\}\,.
\end{align}

Expanding now $V$ of \eqref{potf} in powers of $C \bar{C}$ we obtain 
\begin{align} \label{eq:potexpG}
    V_{\textrm{eff}} = V_0 (T, X, \bar T,\bar X) + C\bar{C}\, V_1 (T, X, \bar T,\bar X) + \mathcal{O} \big( (C \bar C)^2 \big)\,,
\end{align}
where
\begin{align} \label{eq:v0G}
    V_0 &=\frac{ -(X+\bar{X})+2 \big(T \bar{X}+ X \bar{T}\big)- \big[ X \bar{f} (\bar{X})+ \bar{X} f(X) \big] }{ 4\left(T+\bar{T}\right)^2}\,, \\ \label{eq:v1G}
    V_1 &= \frac{f'(X)+ \bar{f}'(\bar{X})}{T+ \bar{T}}\, 2V_0 + \frac{2 - (T+\bar{T}) +f(X) +\bar{f}(\bar{X}) - \big[ X f'(X) +  \bar{X} \bar{f}'(\bar{X})\big]}{\left(T+\bar{T}\right)^2}\,.
\end{align}
To $0$-th order in $C \bar{C}$, the e.\ o.\ m.\ of the superfield $\bar X$ reads
\begin{align} \label{eq:XeomG}
    1-2 T+ f(X)+ X \bar{f}'(\bar X) =0\,,
\end{align}
so that $V_0$ and $V_1$ can be rewritten as:
\begin{align} \label{eq:v0Gafter}
    V_0 &= \frac{  X \bar{X}  \big[ f'(X) + \bar{f}'(\bar X) \big] }{4\left(T+\bar{T}\right)^2} 
\end{align}
and
\begin{align}\label{eq:v1Gafter}
\begin{aligned}
        V_1 &=  \frac{2+f(X)+\bar{f}(\bar{X}) - X \big[\bar{f}'(\bar X) + 2 f'(X) \big] - \bar{X} \big[ f'(X) + 2 \bar{f}'(\bar X)]}{ 2 \left(T+\bar{T}\right)^2} \\  &\quad 
     + \frac{   X\bar{X}  \big[ f'(X) + \bar{f}'(\bar X) \big]^2 }{2 \left(T+\bar{T}\right)^3}     \,.
\end{aligned}
\end{align}
Note that to $0$-th order in $C \bar C$, the potential $V_0$ is positive, due to the metric positivity condition \eqref{positivityCgen}.

Subtracting the e.\ o.\ m.\ of the superfield $\bar X$ from that of $X$ yields
\begin{align} \label{eq:XeomGsub}
    -2T_{\mathtt I} + f_{\mathtt I}( X_{\mathtt R},  X_{\mathtt I}) - X_{\mathtt R} f'_{\mathtt I}( X_{\mathtt R},  X_{\mathtt I})  + X_{\mathtt I} f'_{\mathtt R}( X_{\mathtt R},  X_{\mathtt I})  =0\,,
\end{align}
where $f'_{\mathtt R}$ and $f'_{\mathtt I}$ stand for the real and imaginary parts of the function $\frac{\partial f}{\partial X}$. Moreover, we may Taylor--expand the function $f(X)$ around $X_{\mathtt I}=0\,$ as
\begin{align}
    f(X) =  \sum_{n=0}^\infty \frac{1}{n!}  (\mathrm{i}X_{\mathtt I})^n f^{(n)} (X_{\mathtt R})  \,,
\end{align}
so, after some algebra, we have that
\begin{align} \label{eq:v0ri}
    V_0 =  \frac{ X_{\mathtt R}^2 + X_{\mathtt I}^2}{16\, T_{\mathtt R}^2} \sum_{n=0}^\infty \frac{1}{n!}  \,  (\mathrm{i}X_{\mathtt I})^{n} f^{(n+1)} (X_{\mathtt R}) \, \big[ 1 + (-1)^{n} \big]\,.
\end{align}
It follows that there are no odd powers in $X_{\mathtt I}$, which was expected by the $X\leftrightarrow{\bar X}$ symmetry, while  \eqref{eq:XeomGsub} implies that $X_{\mathtt I}$ is linear in $T_{\mathtt I}$, and thus $X_{\mathtt I}=0=T_{\mathtt I}$ is an extremum of the $0$--order in $C\bar C$ potential \eqref{eq:v0ri}. By focusing next on the quadratic terms in $X_{\mathtt I}$ and imposing positivity of its coefficient, we find the condition
\begin{align}\label{f'''}
  {X_{\textrm{R}}^2}\,  f'''(X_{\textrm{R}})<{2} f'(X_{\textrm{R}})\,,
\end{align}
so that the field $T_{\mathtt I}$ is heavy for any $T_{\mathtt R}$ (in particular during inflation) and can be set to zero:
\begin{align}
X_{\mathtt I} = T_{\mathtt I} = 0\,.
\end{align}

Turning to the next order in $C\bar C$ in the potential \eqref{eq:potexpG}, the first term of the expression \eqref{eq:v1Gafter} for $V_1$ does not exclude the field $C$ becoming tachyonic for large $T_{\mathtt R}$ (while the second term is positive definite). Let us then introduce the dilaton, as in the previous subsections of $R+R^2+R^3$. The K\"ahler derivatives \eqref{eq:kaehlerderG} remain  unchanged and we additionally have
\begin{align}
    D_S W = \frac{\sqrt{3} C \big[ 1-2 T+ f(X) - X f'(X) \big] }{2 \left(S+\bar{S}\right)}\,,
\end{align}
while the K\"ahler metric takes the form \eqref{eq:kaehlerR3dil}, where $\mathbf{K}_{3\times 3}$ is now the K\"ahler metric \eqref{eq:kaehlerG} before the addition of the dilaton. Expanding the new potential $V^{\textrm{dil}}$ around small $C \bar{C}$ as in \eqref{eq:potexpG}, we find that $V_0$ of \eqref{eq:v0G} acquires a prefactor of $1/(S +\bar S)$, while $V_1$ of \eqref{eq:v1G} becomes
\begin{align}
    V_1^{\textrm{dil}} = \frac{V_1}{S+\bar S}  + \frac{3\big|1-2 T+ f(X) - X f'(X)\big|^2  }{4(S+\bar S) \left(T+ \bar{T}\right)^3}\,.
\end{align}
One can now use the equation of the auxiliary field $X$ \eqref{eq:XeomG} to $0$-th order in $C \bar{C}$, which remains unchanged in the presence of the dilaton. It is then straightforward to show that
\begin{align}
    V_1^{\textrm{dil}} = \frac{ 4\big| T + \bar{T} - X \big(f'(X)  + \bar{f}'(\bar X)\big) \big|^2 + X \bar{X} \big(f'(X)  + \bar{f}'(\bar X) \big)^2 }{ 4(S+\bar S) \left(T+ \bar{T}\right)^3}\,,
\end{align}
which is manifestly positive, so that the mass--squared of $C$ becomes again positive in the presence of the dilaton. The tachyonic instability of the goldstino superfield $C$ is thus cured by the dilaton for the general case of $f(R)$ supergravity.

Let us also compare the above results with the bosonic potential \eqref{f(R)potential} of $F(R)$ gravity. To $0$--th order in $C \bar C$, and for $T_I=0$ (real inflaton direction) the potential \eqref{eq:v0G} of $f(R)$ supergravity (coupled to the dilaton)  takes the form
\begin{align} \label{eq:v0Gc}
    V_0 =  \frac{X_{\mathtt R}^2\, f'_{\mathtt R}(X_{\mathtt R})}{\big[1+f_{\mathtt R}(X_{\mathtt R})+X_{\mathtt R} f'_{\mathtt R}(X_{\mathtt R})\big]^2}
\end{align}
where we have used the e.\ o.\ m.\ \eqref{eq:XeomG} of the real field $X_{\mathtt R}$. Comparing the expressions \eqref{f(R)potential} and \eqref{eq:v0Gc} and $\varphi$ in \eqref{eq:f2} with \eqref{eq:XeomG} for $T,X$ real, we thus find that the inflaton and the auxiliary fields are given by the identification
\begin{align}
  \varphi = 2 T_{\mathtt R}\,, \quad \chi=  X_{\mathtt R}\,,
\end{align}
while the relation between the function of $F(R)$ gravity and our function $f$ is given by \eqref{eq:ftoF}.

Note that the point \eqref{eq:min} continues to be a supersymmetric global minimum in flat space, since the e.\ o.\ m. \eqref{eq:XeomG} implies that $X_{\mathtt R}$ vanishes as $X_{\mathtt R}\sim (T_{\mathtt R}-1/2)$, because the function $f(X)$ starts with a term linear in $X$. Thus, inspection of the potential \eqref{eq:v0Gc} shows that $T_{\mathtt R}=1/2$ corresponds to a minimum at zero energy, while one can also verify that the supersymmetric conditions $D_T W =  D_C W  = D_S W =0$ hold.

%%%%%%%%%%%%%%%%%%%%%%%%%%%%%%%
\section{A microscopic theory embedding} \label{sec:string}
%%%%%%%%%%%%%%%%%%%%%%%%%%%%%%%

%%%%%%%%%%%%%%%%%%%%%%%%%%%%%%%
\subsection{A brief review of the ANR model} \label{subsec:su(5)}
%%%%%%%%%%%%%%%%%%%%%%%%%%%%%%%

Here we consider a particular microscopic model \cite{Antoniadis:2020txn, Antoniadis:2021rfm}, that of the flipped $SU(5)\times U(1)$, constructed within the free--fermionic formulation of the heterotic string in four dimensions. It corresponds to a $Z_2\times Z_2$ orbifold that breaks maximal ${\cal N}=4$ supersymmetry to ${\cal N}=1$ and all moduli other than the string dilaton are fixed at gauge or discrete symmetric points arising when all worldsheet degrees of freedom are free fermions. Thus, all string excitations are charged under these symmetries. However, the presence of an anomalous $U(1)$ symmetry, which is a common characteristic of chiral heterotic string constructions, generates a Fayet--Iliopoulos (FI) term and destabilises the vacuum, forcing scalar fields to get vacuum expectation values (VEVs) and drive the theory to a new ``nearby'' supersymmetric vacuum, where several symmetries are broken. Since the FI term is generated at one loop, it introduces a small parameter $\zeta$, so that the new vacuum is calculable in perturbation theory around the original free--fermionic point in powers of $\zeta$. 

The inflaton sector consists of two massless states $y$ and $z$ appropriately chosen, neutral under the observable flipped $SU(5)\times U(1)$ gauge symmetry. In the ``charged'' basis, where all fields but the dilaton $S$ and the two chiral superfields, $y$  (coming from a $Z_2$--twisted sector) and $z$ (coming from the untwisted sector), vanish, the K\"ahler potential and the superpotential read 
\begin{align}\label{eq:kaehlerstr}
 K(z,y,S,\bar z, \bar y,  \bar S) & = -\ln (S+\bar S) -2 \ln (1-|y|^2) -2 \ln \big(1-\tfrac12 |z|^2 \big) \,, \\ \label{eq:superpstr}
 W (z,y) & =  \widetilde{M} \, z y (1- \lambda \, y)\,,
\end{align}
respectively, where  $\widetilde{M}$ is now of the order of $\zeta^5 M_{\textrm{s}}$, with $M_{\textrm{s}} := \frac{1}{\sqrt{2\alpha'}}$ being the string scale, and $\lambda$ is an order one parameter, given in terms of ratios of VEVs and numerical coefficients,  which is in principle calculable. Crucially, for $\lambda=1$ and upon defining $y= \rho \, e^{\mathrm i \theta}$ with $0 \leq \rho <1$, the 
scalar potential has a minimum at $\theta=0$ and $z=0$ along the whole real $\rho$-direction, that reads:
\begin{align}
    V|_{\theta = 0 =z} =  {\widetilde{M}}^2 g_{\textrm{S}}^2 \, \Big( \frac{\rho}{1+ \rho} \Big)^2\,,
\end{align}
where $g_{\textrm{S}}$ is the string coupling: $\langle S\rangle \propto 1/g_s^2$. 

In order to facilitate the comparison with Starobinsky supergravity as defined in \eqref{eq:starsukw}, we change variables to the charged basis using the transformation
\begin{align}\label{TCtoyz}
T=\frac12 \frac{1+y}{1-y}\,,\quad C=\frac{z}{1-y}\,,
\end{align}
that brings the K\"ahler potential and superpotential of \eqref{eq:starsukw} to the form
\begin{align}\label{eq:starsukw2}
 K(z,y,\bar z, \bar y, S, \bar S)  = -\ln (S+\bar S) -3 \ln (1-|y|^2-|z|^2) \,,\quad   W (z,y)  =  M \, z y (1-  y)\,,
\end{align}
where we have also performed a K\"ahler transformation \eqref{eq:kahlertrsf} by a holomorphic function
\begin{align}
    J = -3 \ln (1-y) \,.
\end{align}
Comparing with \eqref{eq:kaehlerstr}, \eqref{eq:superpstr}, the superpotentials coincide for $\lambda=1$, while the K\"ahler potential \eqref{eq:kaehlerstr} for $y$ and $z$ is also of no--scale type symmetric manifold with curvature equal to 3, albeit in a factorised form of two logs instead of one log as in \eqref{eq:starsukw2}.
At the ``fine--tuned'' point $\lambda =1$ for $C=0=z$, the expressions \eqref{eq:kaehlerstr}, \eqref{eq:superpstr} with the redefinition $T_{\mathtt R}=e^\phi$ yield the same scalar potential as \eqref{eq:starsukw2}, namely the Starobinsky potential \eqref{eq:starpot}, up to a different normalisation of the inflaton in the exponent ($e^{-\phi}$ instead of $e^{-\sqrt{2/3}\phi}$ due to the splitting of logs in the K\"ahler potential). Note that in the charged basis, the inflation region of large $T_{\mathtt R}$ corresponds to $y\rightarrow 1$.

On the other hand, the inverse transformation
\begin{align} \label{eq:changeb}
    y = \frac{2T-1}{2T+1}  %\,,\quad z = \frac{2C}{2T+1}\,.
\end{align}
brings \eqref{eq:kaehlerstr}, \eqref{eq:superpstr} to the form:
\begin{align}\label{eq:kaehlerstrTz}
 K(T,z,S, \bar T, \bar z, \bar S)  &= -\ln (S+\bar S) -2 \ln (T+{\bar T})-2 \ln \big(1-\tfrac12 |z|^2 \big) \,,\\
 \quad   W (T,z)  &=  \widetilde{M} \, z \left[\big(T -\tfrac12\big) + (1 -\lambda) \big(T-\tfrac12\big)^2 \right]\,, %\quad  \delta := 1 -\lambda\,,
 \label{eq:prepstrTz}
\end{align}
where we have also performed a K\"ahler transformation \eqref{eq:kahlertrsf} with $J = - 2 \ln (T+\frac{1}{2})$.

%%%%%%%%%%%%%%%%%%%%%%%%%%%%%%%
\subsection{The string embedding} \label{subsec:embed}
%%%%%%%%%%%%%%%%%%%%%%%%%%%%%%%

Now let us compare the scalar potential of the ANR model with the scalar potential of $f(R)$ supergravity constructed in Section~\ref{subsec:f(R)}. The string superpotential \eqref{eq:prepstrTz} and its equivalence for $\lambda=1$ with the Starobinsky supergravity superpotential \eqref{eq:starsukw} described in the previous subsection suggests that for $\lambda\ne 1$ the string scalar potential at $z=0$ should be reproduced by a $f(R)$ supergravity for some particular function $f$.
In fact, in the presence of higher order corrections in $\zeta$ (equivalent to the $\alpha'$--expansion of higher order operators in $W$ involving fields that get VEVs), the superpotential of the ANR model can in general be written as a series expansion in powers of the variable $(T-\tfrac12)$ which corresponds to powers of $y$ in the charged basis\footnote{Note that corrections in higher powers of $z$ are harmless for the potential at $z=0$.}:
\begin{align}
    W(T,z) = z \, g(T) \,, \quad g(T)/{\widetilde M} := 
    \sum_{n\ge 1} g_n (T -\tfrac12)^n\,,
\end{align}
with coefficients $g_1=1$, $g_2=1-\lambda\equiv\delta$ and in general $g_n\sim{\cal O}(\delta^{n-1})$. 

To $0$--order in $z$, the string scalar potential becomes
\begin{align} \label{eq:pot1}
    V_{\textrm{string}} =   \frac{g(T) \bar{g}(\bar T)}{(S+\bar S)(T+\bar T)^2}\,,
\end{align}
while that of $f(R)$ supergravity to $0$--th order in $C$ reads
\begin{align} \label{eq:pot2}
        \frac{1}{S+\bar S}  V_0= \frac{  X \bar{X}  \big[ f'(X) + \bar{f}'(\bar X) \big] }{4 (S+\bar S) \left(T+\bar{T}\right)^2} \,,
\end{align}
following the analysis of Section \ref{subsec:f(R)}. Taking all fields real during inflation, the identification of the potentials \eqref{eq:pot1} and \eqref{eq:pot2} thus imposes
\begin{align} \label{eq:relation}
    g^2(T) \overset{!}{=} \tfrac12 X^2 f'(X)\,,
\end{align}
where we have suppressed the index $\mathtt R$ everywhere. 
Thus \textit{any} function $f(R)$ can be reproduced by a function $g(T)$ defined in \eqref{eq:relation} above, upon solving the e.\ o.\ m.\  \eqref{eq:XeomG} to express the auxiliary field as $X=X(T)$.

The reverse is also true. Differentiating the relation \eqref{eq:relation} with respect to (w.\ r.\ t.) $T$ yields
\begin{align} \label{eq:relation2}
   2\, g(T) \dot{g}(T) = \tfrac12\, X  \big( 2 f'(X) + X f''(X) \big) \frac{\mathrm{d}X}{\mathrm{d}T} \,,
\end{align}
where differentiation w.\ r.\ t.\ $T$ and $X$ are denoted by a dot and prime respectively. Meanwhile, differentiating the e.\ o.\ m.\  \eqref{eq:XeomG} of $X$ w.\ r.\ t.\ $X$ we have that 
\begin{align}
    2 f'(X) +  X f''(X) = 2\frac{\mathrm{d}T}{\mathrm{d}X}\,,
\end{align}
which  can be substituted in the expression \eqref{eq:relation2} to find
\begin{align} \label{eq:res1}
    X= 2 \, g(T) \dot{g}(T)\,.
\end{align}
Combining the result \eqref{eq:res1} with the relation \eqref{eq:relation} we thus have that
\begin{align}
    f'(X) = \frac{1}{2\,\dot{g}^2(T)}
\end{align}
and, using again the  e.\ o.\ m.\  \eqref{eq:XeomG} of $X$ we find that
\begin{align}\label{eq:relationInv}
    f(X) = -\frac{X}{2\dot{g}^2(T)} +2 \big(T-\tfrac12 \big)\,.
\end{align}
This implies that \textit{any} function $g(T)$ can be reproduced by a function $f(X)$, and thus $F(X)$ using the identification \eqref{eq:ftoF}, as defined in \eqref{eq:relationInv} above and upon inverting the equation \eqref{eq:res1} to find $T=T(X)$.

The above equivalence, in a perturbative expansion, identifies the deformation parameter $\delta=1-\lambda$ in the superpotential \eqref{eq:prepstrTz} of the ANR string model with the parameter $\beta$ of the $R^3$ deformation of the Starobinsky model and the mass scales as
\begin{align}\label{deltabeta}
    \delta \simeq - 2\beta\,, \quad \widetilde{M}^2 \simeq 3 m^2  \,,
\end{align}
which can be derived by comparing the string potential \eqref{eq:pot1} with the potential $\frac{m^2}{S+\bar S}  V_0(T_{\mathtt R})$  with $V_0(T_{\mathtt R})$ given by \eqref{eq:v0smallbeta}. Note that if $g(T)$ of the ANR string model is exact, i.e. the superpotential \eqref{eq:prepstrTz} has no higher order in $\alpha'$ corrections, it leads to the same scalar potential as $F(R)$ gravity (modulo the redefinition $T_{\mathtt R}=e^\phi$) starting with $R+ \alpha R^2 + \beta R^3$ with a coefficient $\beta$ given by \eqref{deltabeta}, followed by a series of higher powers $R^{3+n}$ (for $n\ge 1$), with coefficients of order $(1-\lambda)^{1+n}$ determined by the equation \eqref{eq:relationInv}. It follows that since $\beta$ has to be very small $(\lesssim 10^{-4})$ by both experimental constraints and theoretical consistency (see figure~\ref{fig:potR3}), the higher powers of $R$ are extremely suppressed.

%%%%%%%%%%%%%%%%%%%%%%%%%%%%%%%
\section{Conclusions} \label{sec:concl}
%%%%%%%%%%%%%%%%%%%%%%%%%%%%%%%
In this work, we address the problem of the ultraviolet completion of the $R+R^2$ model of Starobinsky inflation. Obtaining it as an effective field theory, one expects corrections in higher powers of the curvature scalar, describing always the same additional to the graviton scalar degree of freedom, the scalaron. One is therefore led to consider a general $F(R)$ gravity. On the other hand, since a fundamental theory of quantum gravity  such as string theory requires supersymmetry at some energy scale, we are led to consider as a first step the embedding of $F(R)$ gravity in supergravity. The final step is to obtain $F(R)$ supergravity as an effective theory within string compactifications.

The main result of this work was first the construction of $F(R)$ $\mathcal{N}=1$ supergravity whose bosonic component contains the $F(R)$ gravity, generalising the $R+R^2$ Starobinsky model of inflation and its supergravity generalisation. We have shown that it can be linearised as ordinary supergravity coupled to two chiral multiplets, consisting of the scalaron $T$, playing the role of the inflaton in Starobisnky inflation, and the goldstino $C$ supermultiplets. The positivity of the $C$ kinetic terms imply a condition on the first derivative of the function $F$, which is the same as the positivity of the scalar potential of $F(R)$ gravity. Besides these two standard multiplets, there is also a third chiral superfield $X$ which is auxiliary around the vacuum of vanishing $C$. Self--consistency then requires that the scalar component of $C$ is not tachyonic along the whole real direction of the inflaton. Indeed, we demonstrated that this condition is automatically satisfied in the presence of the string dilaton when added to the $F(R)$ supergravity action. On the other hand, the pseudoscalar component of the inflaton is also non--tachyonic along the whole real direction of the inflaton scalar component, under an additional condition on the third derivative of $F$. Consequently, it can also be set to zero and the scalar potential reduces to the one of the scalaron of $F(R)$ gravity.
Note that the two conditions we found here on the first and third derivatives of $F$, \eqref{positivityV} and \eqref{f'''}, with $f'(\chi)=(F/\chi)'$, are required for the absence of ghosts and tachyons in $F(R)$ supergravity, in addition to the known conditions \eqref{eq:conditionsf} of the bosonic $F(R)$ gravity, and are associated to the extra degrees of freedom arising in the supersymmetric extension.

As a second step, we identified an embedding of $F(R)$ supergravity within four dimensional heterotic string constructions using the free--fermionic formulation, focusing on a particular model which also has interesting particle physics phenomenology. Although the K\"ahler potentials of $T$ and $C$ are not identical, they are both of no--scale type, symmetric K\"ahler manifolds of curvature equal to 3, while the superpotentials are identical and lead to the same scalar potential of $F(R)$ gravity. However, the string model uses a different parametrisation in terms of a different analytic function $g(T)$. In this work, we have demonstrated a one--to--one correspondence between $F(R)$ and $g(T)$ that allows to obtain one function in terms of the other in both directions. 

A physically interesting case is to consider $F(R)$ as a small deformation of the Starobinsky model of inflation by the addition of higher powers of $R$, starting with an $R^3$ term. When its coefficient is negative and small, the scalar potential rises sharply in the direction of large inflaton values, restricting its excursion within a finite range, below the mass scale of the KK tower where EFT validity breaks down, as dictated by the Swampland distance conjecture. Thus, a sufficiently small deformation solves the problem of initial conditions in the Starobinsky model, which is of typical large field type. Moreover, this deformation improves agreement with recent ACT cosmological data.

A remaining open problem is the implementation of a string dilaton stabilisation mechansim which also fixes the string coupling and thus the value of the FI term that controls the perturbative expansion parameter away from the free--fermionic vacuum. It would also be interesting to investigate a possible holographic description of $F(R)$ supergravity in the context of \cite{Skenderis:2007sm,Bzowski:2012ih}.

\section*{Acknowledgements}

We would like to thank Keith A.~Olive and Augusto Sagnotti for enlightening discussions and the Center for Cosmology and Particle Physics at New York University, where part of this work was performed, for hospitality. This research of I.A. was supported in part by the Higher Education and Science Committee of MESCS RA (Research Project N 24RL-1C036). C.M. thanks Kostas Skenderis for a useful discussion, gratefully acknowledges support from the Simons Center for Geometry and Physics, Stony Brook University during the programme ``50 years of the black hole information paradox'' and from the Institute for Advanced Study at Princeton, at which some of the research for this paper was performed, and is supported by a fellowship of the Scuola Normale Superiore and by INFN (I.S. GSS-Pi).

\appendix
%%%%%%%%%%%%%%%%%%%%%%%%%%%%%%%
\section{\texorpdfstring{Elements of \ensuremath{\mathcal{N}=1} supergravity}{Elements of N=1 supergravity}} \label{app:ids}
%%%%%%%%%%%%%%%%%%%%%%%%%%%%%%%

Here we gather elements of $\mathcal{N}=1$ supergravity following \cite{Wess:1992cp, Cremmer:1978hn, Kugo:1982cu, Cecotti:1987sa, Butter:2009cp, Freedman:2012zz}. The vielbein chiral density $\mathcal{E}$ contains the vielbein determinant $e$ in its lowest component, while the chiral curvature superfield $\mathcal{R}$ contains the Ricci scalar $R$ in its \textit{upper} component,  namely
\begin{align} \label{eq:conventions}
\mathcal{E} \big | = e = \sqrt{-g}\,, \quad     \mathcal{R} \big|_{\Theta^2} = -\tfrac16 R\,,
\end{align}
where we display only the bosonic (and non--auxiliary) components of $ \mathcal{R}$. Note that our conventions slightly differ from those of \cite{Wess:1992cp}. The superfield $\mathcal{R}$ can also be defined via the action of the chiral projector $\mathscr{P}$ on the chiral compensator superfield $S_0$ as
\begin{align}\label{eq:defr}
    \mathcal{R} := \frac{\mathscr{P}(\bar{S}_0)}{S_0} \,.
\end{align}
The chiral projector $\mathscr{P}$ is the analogue of $\bar{D}^2$  of global supersymmetry and is defined as
\begin{align}
\mathscr{P} := -\tfrac14(\bar{\mathcal{D}}^2 - 4 \mathcal{R})\,,
\end{align}
where $\mathcal{D}_\alpha$ is the covariant derivative in curved superspace. Just like the operator $\mathcal{D}^2$ selects the highest component of a chiral superfield $\Phi$,
\begin{align}
\Phi \big|_{\Theta^2} = F \quad \Rightarrow  \quad   -\tfrac14 {\mathcal D}^2 \Phi \big| =  F \,,
\end{align}
we also have that
\begin{align} \label{eq:curvconv}
    \mathcal{R} \big|_{\Theta^2} = -\tfrac16 R \quad \Rightarrow  \quad \mathscr{P}( \bar{\mathcal{R}} ) \big| =  - \tfrac16  R \,,
\end{align}
namely the Ricci scalar $R$ appears in the \textit{lowest} component of the superfield $\mathscr{P}(\bar { \mathcal{R} }) $ respectively.

The Lagrangian of $\mathcal{N}=1$ supergravity coupled to $N$ chiral multiplets $\Phi^I$, $I=1,\ldots,N$, can be written as
\begin{align}\label{eq:supergr2}
 \mathcal{L}  = -3 \Big[ e^{-K(\Phi^I,\bar{\Phi}^{\bar J})/3}    \Big]_\D + \Big( \Big[   W\big(\Phi^I\big)  \Big]_\F+ \textrm{h.c.}\Big)\,,
\end{align}
where $K(\Phi^I,\bar{\Phi}^{\bar J})$ and $W(\Phi^I)$ are the K\"ahler potential and superpotential respectively, or as
\begin{align}\label{eq:superconfl}
  \mathcal{L}  = -3 \Big[ e^{-K(\Phi^I,\bar{\Phi}^{\bar J})/3}   S_0 \bar{S}_0 \Big]_\D + \Big( \Big[   W\big(\Phi^I\big) S_0^3 \Big]_\F+ \textrm{h.c.}\Big)\,,
\end{align}
with the gauge--fixing of the compensator
\begin{align} \label{eq:gauge}
S_0 =1  = \bar{S}_0
\end{align}
yielding back the form \eqref{eq:supergr2}. Let us also recall a useful identity. The \D\, density of any function $\mathcal{H}(\Phi,\bar \Phi)$ can be written as an \F\, density and vice versa via
\begin{align} \label{eq:ids}
\big[ \mathcal{H}(\Phi,\bar \Phi) + \textrm{h.c.} \big]_\D = \Big[ \mathscr{P}\big( \mathcal{H}(\Phi,\bar \Phi) \big)  \Big]_\F + \textrm{h.c.}
\end{align}
For example, for a chiral superfield $T$, it holds that \cite{Cecotti:1987sa,Ferrara:2013wka}
\begin{align} \label{eq:id1}
 (T+\bar T)S_0 \bar{S}_0\big|_\D = TS_0 \mathscr{P}(\bar{S}_0)  \big|_\F  + \textrm{h.c.} = T \mathcal{R} S_0^2 \big|_\F  + \textrm{h.c.}   
\end{align}

Finally, after the gauge--fixing \eqref{eq:gauge} that takes gravity kinetic terms into the Einstein frame, the scalar potential of matter--coupled $\mathcal{N}=1$ supergravity \eqref{eq:superconfl} reads
\begin{align} \label{eq:potgens}
    V = e^K(g^{I\bar J} D_I W D_{\bar J} \bar W  - 3|W|^2)\,,
\end{align}
where $g_{I \bar J}:= K_{I \bar J} := \partial_I \partial_{\bar J} K $ is the K\"ahler metric and the covariant derivative acts as $D_I W :=\partial_I W + K_I W $, with $K_I := \partial_I K$. Let us also recall that the scalar potential \eqref{eq:potgens} is invariant under  \textit{K\"ahler transformations},
\begin{align} \label{eq:kahlertrsf}
    \begin{aligned}
        K \big(\Phi^I ,\bar{\Phi}^{\bar{J}} \big) &\rightarrow  K \big(\Phi^I ,\bar{\Phi}^{\bar{J}} \big)  + J \big(\Phi^I\big) + \bar{J}\big(\bar{\Phi}^{\bar{J}}\big)\\
        W  (\Phi^I ) & \rightarrow e^{-J(\Phi^I )}  W (\Phi^I ) \,,
    \end{aligned}
\end{align}
where $J \big(\Phi^I\big) $ is any holomorphic function of the chiral superfields $\Phi^I $.

\newpage
%%%%%%%%%%%%%%%%%%%%%%%%%%%
\bibliographystyle{utphys}

\providecommand{\href}[2]{#2}\begingroup\raggedright\endgroup

\end{document}